\documentclass[11pt]{article}

\usepackage{epsfig}
\usepackage{amsmath}
\usepackage{amsfonts}
\usepackage{amssymb}

\def\rd{{\rm d}}

\def\bea{\begin{eqnarray}}
\def\eea{\end{eqnarray}}
\def\beq{\begin{equation}}
\def\eeq{\end{equation}}

\def\be{\begin{equation}}
\def\ee{\end{equation}}

\def\5{\overline 5}
\def\vp{\varphi}

\def \wt {\widetilde}

\newcommand{\vol}{d^4 x \sqrt{- g}}
\newcommand{\volt}{d^4 x \sqrt{- \widetilde g}}

\def\beq{\begin{equation}}
\def\eeq{\end{equation}}
\newcommand{\gsim}{\mbox{\raisebox{-1.ex}{$\stackrel
     {\textstyle>}{\textstyle\sim}$}}}
\newcommand{\lsim}{\mbox{\raisebox{-1.ex}{$\stackrel
     {\textstyle<}{\textstyle \sim}$}}}

\begin{document}
\begin{titlepage}

\begin{flushright}
\vspace{-2cm}
Bicocca-FT-04-4
\end{flushright}

\begin{center}

\begin{center}

\vspace{1.7cm}

{\LARGE\bf Dilatonic ghost condensate as dark energy}

\end{center}

\vspace{1.4cm}

{\Large Federico Piazza$^1$ and Shinji Tsujikawa$^2$}

\vspace{1.2cm}

$^1$Dipartimento di Fisica and INFN, Universit\`a di Milano 
Bicocca, \\ Piazza delle Scienze 3, I-20126 Milano, Italia

\vspace{0.3cm}

$^2$Department of Physics, Gunma National College of 
Technology, 580 Toriba, \\
Maebashi, Gunma 371-8530, 
Japan 

\end{center}

\vspace{0.8cm}

\begin{abstract}
We explore a dark energy model with a 
ghost scalar field in the context of 
the runaway dilaton scenario 
in low-energy effective string theory. 
We address the problem of vacuum stability
by implementing
higher-order derivative terms and show that 
a cosmologically viable model of ``phantomized'' dark energy 
can be constructed without violating the stability of quantum
fluctuations. We also analytically derive the condition 
under which cosmological scaling solutions exist
starting from a general Lagrangian including the phantom
type scalar field.
We apply this method to the case where the dilaton 
is coupled to non-relativistic dark matter  and find that 
the system tends to become quantum mechanically 
unstable when a constant coupling is always present.
Nevertheless, it is possible to obtain a viable cosmological 
solution in which the energy density of the 
dilaton eventually approaches the 
present value of dark energy provided that the coupling 
rapidly grows during the transition to the scalar field
dominated era.
\end{abstract}

\end{titlepage}

\section{Introduction}                            

One of the most challenging and intriguing problems 
of cosmology is undoubtedly that of dark energy, the marginally 
dominant negative pressure component that drives the present acceleration 
of the Universe (for reviews see e.g. \cite{varun,paddy}). 
The identification between dark energy and the energy of the vacuum 
(cosmological constant),
which, in some respect, may look as a ``minimal'' choice,
rises at least a couple of embarrassing issues. The main concern is perhaps
to explain why the vacuum energy 
is so small in particle physics units (``fine-tuning'' problem). Moreover, 
beside being generically ``small'', the cosmological constant
happens to be exactly of the value required to become dominant 
at the present 
epoch. The latter mysterious circumstance is sometimes
referred to as the ``coincidence problem''.

These puzzles may be better interpreted by assuming that the energy of the 
vacuum is, 
for some unknown cancellation mechanism, exactly zero and by considering
in its place a dark energy component with a dynamically variable 
equation of state. A host of such models have been studied, ranging from 
quintessence \cite{quin}, K-essence \cite{kesse},
braneworlds \cite{brane}, tachyons \cite{tachyon}, 
chaplygin gas \cite{chap} etc..
Effective scalar fields represent, in this respect, a simple and well 
motivated choice, since they are omnipresent in supersymmetric field 
theories and in string/M theory. 
For instance, a definite prediction of string theory is that the gauge and 
gravitational couplings are not fixed \emph{a priori} 
but, rather, related to the vacuum expectation value of a scalar field, 
the dilaton. More precisely, at the tree level in the
string loop expansion, the vacuum expectation value 
of the (four dimensional) dilaton $\phi$ is related to the gauge 
coupling $\alpha_{\rm GUT}$  and to the Planck mass $M_P$ 
through $e^\phi \simeq M_s^2/M_P^2 \simeq \alpha_{\rm GUT}$, where
$M_s\equiv \sqrt{2/\alpha '}$ is the string mass and $\alpha '$ the 
universal 
Regge slope parameter of the string. 

The very existence of a relation between the couplings and the 
vacuum expectation value (VEV) of
a scalar field, although promisingly tasting of ``unification'', threatens
with possible violations of the equivalence principle as well as
unobserved variations of the coupling constants.
It is common wisdom \cite{TV88} to assume that the dilaton and  
the other moduli of the theory
are given a mass by some non-perturbative mechanism, in such a way 
that their long range interactions are suppressed and their VEVs frozen at
some phenomenologically reasonable value. 
An alternative possibility is that, at the level of the effective action,
the dilaton decouples from the other fields. To be more explicit, 
let us consider a generic effective string action at lowest order in 
$\alpha '$ [we use the convention $(- + + +)$ for the metric]
\begin{equation} \label{intro}
{\cal S} = \frac{1}{\alpha'} \int \volt \, \left[B_g(\phi) \wt R  + B_\phi(\phi)\, 
\wt{g}^{\, \mu\nu} \, \partial_\mu  \phi \, \partial_\nu \phi \, - 
\frac{\alpha'}{4} B_F(\phi){\wt F}^2 + \, \dots \, \right]\,, 
\end{equation} 
where $\wt R$ is the scalar curvature and ${\wt F}$ is the gauge field.
Here the dilaton-dependent loop effects as well as the non-perturbative 
corrections are encoded in the ``coupling functions'' $B_i(\phi)$, and 
a tilde denotes the quantities as measured in the ``string frame'', the 
sigma-model metric minimally coupled to the fundamental strings. 

In the non--weak coupling region ($e^\phi~\gsim~1$), where the loop 
effects are important, all the (relevant)
functions $B_i(\phi)$ may extremize at some value $\phi_m$. 
This possibility has been investigated in Ref. \cite{DP94} where it has 
been shown that the cosmological evolution tends to push the dilaton toward the 
value $\phi_m$. This is a phenomenologically ``safe'' vacuum where the 
massless
dilaton decouples from the matter fields.

By means of a large-N argument, Veneziano has recently suggested \cite{V01}
that the effective couplings $B_i$ are, for the most part, 
\emph{induced} by the quantum corrections of the many moduli and gauge 
bosons of the theory. As a result, these functions of $\phi$ reach 
an extremum at infinite \emph{bare} coupling $e^\phi \rightarrow \infty$ 
and
follow the general behaviour
\begin{equation}
\label{Bidef}
B_i(\phi) = C_i + {\cal O}(e^{-\phi}) ~~~~~~~
(e^\phi \gg 1).
\end{equation}
In this scenario  (of the so called ``runaway dilaton''), $\phi$ gradually 
decouples 
from gravity and from the matter fields by evolving towards infinity: 
$\phi_m = \infty$.
The moduli of the coefficients  $C_i$ are proportional, by factors of 
order one, to the number $N\sim 10^2$ of independent degrees of 
freedom which have been integrated over. What about their signs? 
It is natural to require a sensible low energy theory e.g. -- in our 
conventions -- that both $C_g$ and $C_F$ be positive numbers.
Moreover, in order for the dilaton $\phi$ to behave as a 
``normal'' scalar field in the limit of $\phi \gg 1$,
the kinetic coefficient $C_\phi$ has to be negative. 

This choice was made in 
\cite{GPV,DPV1,DPV2} where the late-time cosmology and 
phenomenology of this model with a ``well-behaved'' scalar field were 
studied in some detail. In particular, in \cite{GPV} it was shown that 
a residual coupling of the dilaton to some hidden ( --dark matter?) 
sector of 
the theory can give rise to a final cosmological attractor with both  
an accelerated expansion
and a constant ratio between dark matter and dilatonic energy 
densities. 
Such a type of ``scaling'' behaviour \cite{wetter,Amendola2}, relaxes
the ``coincidence problem'' by interpreting the present energy
budget of the Universe (dark matter $\simeq 1/3$, dark energy $\simeq 2/3$ 
of the total) as belonging to a stable attractor configuration. 
Moreover, these scaling solutions can lead to a viable 
late-time 
cosmology \cite{AGUT} with the acceleration starting earlier ($z>1$) than 
in usual 
(uncoupled) dark energy models but still consistent with the recent
Type Ia supernovae data \cite{Amendola3,AGPV}. 

The opposite choice for the kinetic sign in (\ref{intro}),
$C_\phi >0$, leads to what in quantum field theory is 
called a ``ghost'' field or, in a more recent cosmological fashion, a 
``phantom'' 
\cite{caldwell}. 
When considered as dark energy candidates, ghosts can have an equation of 
state parameter
$w\equiv p/\rho< -1$ which is not ruled out (rather, slightly favoured 
\cite{tirthalam}) by current 
observations. Ghosts/phantoms prove to have a viable cosmological behavior 
which has been extensively studied in 
Refs.~\cite{caldwell,phantom,Sami:2003xv} and 
also constrained with observations \cite{sami}. 
Although intriguing as ``classical -- cosmological'' fields, phantoms 
are generally plagued by severe UV quantum instabilities\footnote{The 
energy of a phantom field is 
not bounded from below and this makes the vacuum unstable against the 
production of ghosts + 
normal (positive energy) fields \cite{Carroll}. 
Even when decoupled from the matter fields, ghosts couple to gravitons 
which mediate
vacuum decay processes of the type $vacuum \rightarrow 2\, ghosts + 2 
\gamma$ 
and an unnatural Lorenz invariance breaking -- cut off of $\sim$ MeV is 
required to prevent an 
overproduction of cosmic gamma rays \cite{Cline}.}, so that the 
fundamental origin
of a $w<-1$ component still represents an interesting challenge for 
theoreticians.

Recently, it has been shown  that a scalar field with a wrong sign 
kinetic term does not necessarily lead to inconsistencies, provided a 
suitable structure of higher order kinetic terms exists in the effective 
theory \cite{Arkani}. The basic mechanism resembles that of
a $\lambda \vp^4$--theory, where the field acquires a non-zero VEV and the 
energy of the small fluctuations around the new minimum is bounded from 
below.
By the same token, in a theory with higher 
order kinetic terms, the field can ``condensate'' at a 
non-constant background value, $\dot{\vp}_0 \neq 0$, which is perfectly 
stable at the quantum level. 
The basic scheme is that of a Lagrangian of the type 
$p = - \partial \vp^2 + \partial \vp^4/m^4$ where $\vp$ is a canonically 
normalized (ghost) scalar field and $m$ a mass scale. In this case a 
background
value $\vp_0$ can, at the same time, be quantum mechanically stable 
($\partial \vp_0^2\geq m^4/2$) and act as a negative pressure 
component ($\partial \vp_0^2\leq m^4$). 
A kinetically driven cosmic acceleration has been proposed both in 
inflationary 
\cite{kinflation,Arkani2} and quintessence \cite{kesse} contexts.
See Refs.~\cite{ghostrecent} for recent works about ghost 
condensation.

In this paper we highlight the runaway dilaton scenario by
exploiting the possibility that string--loop corrections may result in
the ``wrong'' kinetic sign for the dilaton at the effective level
[i.e. $C_\phi >0$ in Eq.~(\ref{Bidef})].
The low energy theory can still be consistent if the dilaton, by a 
mechanism similar to that in \cite{Arkani}, 
``condensates'' in a ``safe'' vacuum configuration. 
We also show that a residual coupling of the scalar field to dark matter 
can lead to a 
late-time accelerating Universe with the fractional densities of 
dark matter and dark energy that remain constant in time.
In section 2 we sketch the origin of our model 
from the low-energy limit of string theory. 
In section 3 we  address the stability of quantum fluctuations for a 
scalar field $\vp$ whose Lagrangian is a general function of 
$\partial \vp^2$ and of $\vp$ itself. We also generalize the late-time attractors studied in 
\cite{wetter,Amendola2,GPV} and derive the general functional form of 
the Lagrangians that allow this type of scaling solutions.
In section 4 we study the dynamics of the effective ghost/dilaton
for various choices of the parameters.
It is found that in the absence of the coupling $Q$ between the dilaton
and the non-relativistic matter one can get a viable attractor solution
which asymptotically approaches a de Sitter phase
without violating the stability of the vacuum.
We also show that there exist well-behaved cosmological scaling solutions
if the coupling $Q$ grows from nearly zero to a constant 
value during the transition to the dilaton-dominant era.
We give conclusions and discussions in the final section.

\section{A string-inspired model}

Let us  consider a general four-dimensional effective 
low-energy string action \cite{Green}
\begin{eqnarray} 
\label{action}
{\cal S} = \frac{1}{\alpha'} \int \volt \, \Big[B_g(\phi) \wt R  + 
B_\phi(\phi)\, 
\wt{g}^{\, \mu\nu} \, \partial_\mu  \phi \, \partial_\nu \phi \, 
- \alpha' \wt V(\phi) 
+\ \dots \qquad \qquad \qquad \nonumber \\
+ \ \ {\rm higher\ order \ in\ } \alpha'  \ \Big] \, +\, \wt{{\cal 
S}}_m[\phi, 
\wt{g}_{\mu\nu}, \Psi_i]\,,
\end{eqnarray}
where $\wt R$ is the scalar curvature and 
$\phi$ is the (four dimensional) dilaton field with potential $\wt 
V(\phi)$.
Here $\wt{\cal{S}}_m$ is the action for the matter fields $\Psi_i$, 
which we suppose that they are generally coupled to the dilaton.
The dilaton-dependent loop effects as well as the non-perturbative 
corrections are encoded, at the lowest order of approximation in the 
universal 
Regge slope parameter $\alpha'$, in the ``coupling functions'' 
$B_i(\phi)$. 
In the weak coupling region of the theory, where the tree-level  
(four dimensional) string coupling $g_s^2 \equiv e^\phi$ is much smaller 
than one,  the functions $B_i(\phi)$ can be expressed by an expansion
of the form 
\begin{equation} \label{expansion}
B_i(\phi) = e^{-\phi} + c_0 + c_1 e^\phi + \dots~~~~~
(e^\phi \ll 1). 
\end{equation}

As argued in Ref.~\cite{DPV1}, we will assume that in the region 
with $\phi \gg 1$, the coupling functions $B_i$ have 
the following behavior
\begin{equation}\label{triv}
B_i(\phi) = C_i + D_ie^{-\phi} + {\cal O}(e^{-2\phi}),
\end{equation}
asymptotically approaching a constant value as $\phi \to \infty$.
Therefore the dilaton gradually decouples from gravity as the field
evolves toward the region with $\phi \gg 1$ from the weakly coupled 
regime. In this scenario, the coefficients $C_i$ and $D_i$ are of order 
$10^2$ (the number of independent degrees of freedom which have been 
integrated over) and unity respectively.

It is convenient to introduce the conformally 
related ``Einstein'' metric \cite{GPV}
\begin{equation}
g_{\mu \nu}\,  = C_g^{-1} B_g(\phi) \wt{g}_{\mu\nu}.
\end{equation} 
In the above relation the constant factor $C_g^{-1}$ has been chosen so 
that the Einstein metric 
approximates the ``string'' one in the limit 
$\phi\rightarrow \infty$. Note that in the runaway dilaton scenario 
\eqref{triv}
the 4-dimensional Planck mass $M_P$ is asymptotically related to
the string scale $M_s \equiv \sqrt{2/\alpha'}$
through $M_P^2 \simeq C_g M_s^2$, 
which makes sense if one considers that 
the coefficients $C_i$ are expected to be of order $10^2$. 
The action in the Einstein frame reads
\begin{equation} 
\label{Eaction}
{\cal S}_E = \int \vol \, \left[\frac{M_P^2}{2} R - \frac{\epsilon}{2}
(\partial \vp)^2  - V(\vp) + {\rm higher\ order \ terms\ } \right]  
+\, {\cal S}_m[\vp, g_{\mu\nu},\Psi_i]\,,
\end{equation}
where we have introduced a canonically defined scalar field $\vp$ with the 
dimensions of a mass and the 
Einstein frame potential $V(\vp)$ as follows
\begin{eqnarray} 
\label{FVdef}
M_P^2\left[\frac{3}{2}
\left(\frac{B_g'}{B_g}\right)^2 - \frac{B_\phi}{B_g} \right] d\phi^2 = 
\epsilon\, d\vp^2\,,
\qquad V(\vp)=C_g^2B_g^{-2}\wt V(\phi)\,.
\end{eqnarray}
Here a prime denotes a derivative with respect to $\phi$ and $\epsilon$ 
evaluates $\pm 1$.  
When the expression in square brackets is positive the dilaton 
behaves as a normal scalar field ($\epsilon = +1$).
For example it is known that at tree level, despite the positive sign of 
the kinetic term in the string-frame ($B_\phi =B_g=e^{-\phi}$), we have, 
in the Einstein-frame, a ``proper'' scalar field.

In the runaway scenario that we are considering (\ref{triv})
the value of $\epsilon$ crucially
depends on the sign of $C_\phi$, since in the $\phi \gg 1$ limit the 
expression in 
square brackets in (\ref{FVdef}) evaluates $-C_\phi / C_g + {\cal 
O}(e^{-\phi})$.
It is not difficult to imagine that loop corrections give the ``wrong'' 
kinetic 
sign to the dilaton, i.e. $C_\phi>0$. This possibility, in principle, 
doesn't worry us too much 
since, as sketched in the introduction and analyzed in some detail in 
section 3, higher order kinetic terms may
ensure the stability of quantum fluctuations 
by ``condensating'' \cite{Arkani}
the ghost field in a stable vacuum. However, if we are to 
interpret a ghost condensate as dark energy, 
a ``fine-tuning'' problem, identical to that of the 
cosmological constant, arises. 
The higher order kinetic terms are in fact generically 
suppressed by powers of the cut-off (string-) scale and therefore are 
expected 
to be largely subdominant today. 
A possible way out, again, is to consider the effects of the loop 
corrections. 
In particular, the contributions of the lightest fields (neutrinos?)
should dominate the higher order corrections giving 
effective terms of the type $(\partial \vp)^{2n}/m^{4 (n-1)}$, $m$ being 
the mass 
of the lightest (coupled) field running into the loops. 

In what follows we will concentrate on an (admittedly, ad-hoc) higher 
order correction to insert in (\ref{Eaction}) of the type (see also 
\cite{Cartier,Tsuji})
\begin{equation}\label{explode}
{\rm higher\ order\ terms\ } = 
\ \frac{A}{m^4} (\partial \vp)^4 e^{\lambda \vp /M_P}\,,
\end{equation}
where $A$ and $\lambda$ are numbers of order one and $m$ a mass scale.  
Exponential corrections of this type are naively expected if the
theory contains ``exponentially light'' ($m \simeq C_m + D_m e^{-\phi} =  
m_0 e^{-\phi}$; i.e. $C_m = 0$) fields running into the loops. 

While considering as our basic structure the one with 
$(\partial \vp)^2 + (\partial \vp)^4$,
for the sake of generality we also cite the possibility of a ``zeroth
order'' --kinetic contribution: a potential term $V$ for the dilaton.
Since there is no trace of it at tree level, $\wt V$ 
must be of non-perturbative origin 
(e.g., $\sim e^{-1/g_s^2}$) and dying out exponentially at weak coupling
($\phi \rightarrow -\infty$). We will make also the ``phenomenological'' 
request that
it goes to zero in the limit $\phi \rightarrow \infty$. A viable ansatz is 
\cite{GPV}
\begin{equation}
\label{po}
\wt V(\phi)=M_0^4  \left[\exp(-e^{-\phi}/\beta_1)-
\exp(-e^{-\phi}/\beta_2)\right]\,,
\end{equation}
where $M_0$ is some mass scale, and $\beta_1$ and $\beta_2$ 
are constants satisfying $0<\beta_2<\beta_1$.
This is a bell-type potential which has a maximum in an 
intermediate regime.
Note that in the strongly coupled region ($\phi \gg 1$) 
the potential behaves as
\begin{eqnarray}
V(\phi) \simeq \left(\frac{1}{\beta_2}-\frac{1}{\beta_1}
\right) M_0^4  e^{- \phi}\,.
\label{Vapp}
\end{eqnarray}

In summary, the most general scalar field Lagrangian that we consider is
\begin{equation} 
\label{lagrangian}
{\cal L}_{\rm dilaton}= \frac{1}{2} (\partial \vp)^2 + 
\frac{A}{m^4} (\partial \vp)^4 e^{\lambda \vp /M_P} 
-\left(\frac{1}{\beta_2} -\frac{1}{\beta_1}\right) 
M_0^4 e^{-\lambda_1 \vp/M_P},
\end{equation}
where, by equation (\ref{FVdef}), $\lambda_1 \simeq \sqrt{C_g/C_\phi}$ .
Before going into the 
details of our model, let us have a look at some general properties of a 
scalar field $\varphi$ whose Lagrangian density is a generic function of 
$(\partial\varphi)^2$ and of $\varphi$ itself.

\section{A (sufficiently) general cosmological scalar field}

In this section we analyze some general features of a canonical
cosmological scalar field $\varphi$
whose lagrangian is a general function of the scalar quadratic kinetic 
term $\partial \varphi^2$ 
and of the field itself \cite{kesse,kinflation}. We consider the following 
action written in the ``Einstein frame'' 
\begin{equation}
\label{lag2}
{\cal S} \ = \ {\cal S}_{grav}+ {\cal S}_\vp + {\cal S}_m \ = \
\int d^4 x \sqrt{-g} \left[\frac{M_P^2}{2}\, R 
+ p(X, \varphi)\right] 
+ {\cal S}_m [\varphi, \Psi_i, g_{\mu \nu}]\,,
\end{equation}
where $X$ is defined as $X\equiv -g^{\mu\nu} \partial_\mu \varphi 
\partial_\nu \varphi /2$, and 
${\cal S}_m$ is the action for the matter fields, which is
generally dependent on $\vp$ as well. 
Hereafter we will use the unit $M_P=1$, 
but we restore the Planck mass when it is needed.
The Lagrangian (\ref{lagrangian}) can be expressed as
\begin{equation}
\label{ourlag}
p(X, \varphi)=-X+c_1e^{\lambda \varphi}X^2-c_2e^{-\lambda_1 \varphi}\,,
\end{equation}
where $c_1=4A/m^4$ and 
$c_2=\left(1/\beta_2 -1/\beta_1\right) M_0^4$.

The energy momentum tensor of the scalar field that we obtain from 
Eq.~(\ref{lag2}) reads
\begin{equation} \label{emtensor}
T^{(\vp)}_{\mu \nu} \ = \
- \frac{2}{\sqrt{-g}} \frac{\delta {\cal S}_\vp}
{\delta g^{\mu \nu}} \ = \ 
g_{\mu \nu} p + p_X \, \partial_\mu 
\vp \partial_\nu \vp\, .
\end{equation}
Here and in the following a suffix $X$ or $\vp$ indicates a 
partial derivative with respect to $X$ and $\vp$ respectively.  
We note that the energy momentum tensor (\ref{emtensor}) of the scalar 
field
is that of a perfect fluid, 
$T_{\mu \nu} = (\rho + p)u_\mu u_\nu + g_{\mu \nu} p$, with velocity 
$u_\mu = \partial_\mu \vp/\sqrt{2X}$ and energy density 
\begin{equation}
\label{rhophi}
\rho = 2 X p_X - p\,.
\end{equation}

%
\subsection{Consistency conditions}

In order to address the problem of the quantum stability of the scalar 
field 
in action (\ref{lag2}) we consider the dynamics of the small 
fluctuations $\delta\vp$ around a background value $\vp_0$ which
is a solution of the equations of motions. Since $p$ depends both on $\vp$ 
and 
its derivatives, the most general case to be
considered is that of a background with both $\vp_0 \neq 0$ and
$\dot{\vp}_0 \neq 0$. In particular, the choice of a
background field evolving in time is essential for the mechanism 
of ``ghost condensation'' \cite{Arkani}, by which a 
scalar theory with the ``wrong'' kinetic sign settles to a stable vacuum.

We divide the field $\vp$ into a homogeneous part
$\vp_0$ and a fluctuation $\delta \vp$, as 
\begin{eqnarray} \label{decompose}
\vp(t, {\bf x})=\vp_0(t)+\delta \vp(t, {\bf x}) \,.
\label{phide}
\end{eqnarray}
Since we are most concerned with UV instabilities it is not restrictive to 
consider a Minkowski background metric and to choose, at least locally, a 
time direction in such a way to satisfy (\ref{decompose}).
By expanding $p(X,\vp)$ at the second order in $\delta \vp$ it is 
straightforward to find the Lagrangian and then the Hamiltonian for the 
fluctuations. The Hamiltonian reads 
\begin{eqnarray}
{\cal H} = \left(p_X +
2 X p_{XX} \right) 
\frac{(\delta \dot{\vp})^2}{2}
+p_X \frac{(\nabla \delta \vp)^2}{2} 
-p_{\vp \vp}  \frac{(\delta \vp)^2}{2}\,.
\label{Ldelpsi}
\end{eqnarray}
The fluctuations of a time-varying scalar typically obey 
Lorentz-violating dispersion relations \cite{Kos}. 
This should not worry us too much since, 
in a cosmological setup, Lorenz invariance is always violated by the
presence of a ``preferred'' (CMB--, comoving observers--) frame.
The positive definiteness\footnote{Of course the overall 
sign of the Hamiltonian for 
$\delta \vp$ has a meaning if compared with 
the sign of the Hamiltonian of 
the other fields, which we take to be positive.} of the Hamiltonian is 
guaranteed if the following 
conditions hold 
\begin{eqnarray}
\label{xi}
\xi_1\,  \equiv \, p_X +
2 X p_{XX} \, \ge \, 0,\qquad 
\xi_2 \, \equiv \,   p_X \, \ge \, 0, \\
 \xi_3 \, \equiv \, -p_{\vp\vp} \, \ge \, 0\,. \label{xi2}
\end{eqnarray}

Both $\xi_1$ and $\xi_2$ are usually taken
into account when considering the stability of classical perturbations.
The quantity often used is the speed of sound $c_s$
defined by \cite{kper} 
\begin{eqnarray}
\label{sound}
c_s^2 \equiv \frac{p_{X}}{\rho_{X}}
=\frac{\xi_2}{\xi_1}\,,
\end{eqnarray}
which in cosmological perturbations appears as a coefficient of the 
$k^2/a^2$ term
(here $a$ is the scale factor and $k$ is the comoving  momentum).
While the classical fluctuations may be regarded to be
stable when $c_s^2$ is positive, it is essential, for the quantum 
stability, 
that both $\xi_1>0$ and $\xi_2 \ge 0$. 
These two conditions, in fact, prevent an instability of a very bad type 
related to the presence of negative energy ``ghost'' states which render the 
vacuum 
unstable under a 
catastrophic production of ghosts and photons pairs \cite{Carroll,Cline}.
This is essentially a UV instability,
since the rate of production from the vacuum 
is simply proportional to the phase space integral
on all possible final states and since only a UV cut off can prevent the 
creation of modes of arbitrarily high energies.
When the Lagrangian $p(X, \vp)$ is written as
$p(X, \vp)=-X-V(\vp)$, as was considered in most 
of the phantom dark energy models \cite{phantom},  
the coefficients $\xi_1$ and $\xi_2$ are both negative.

The situation is different if we take into account 
higher-order terms such as $X^2$ in $p(X, \vp)$.
For example, consider the case with $p=-X+X^2$ \cite{Arkani}. 
Then one has $\xi_1=-1+6X$ and $\xi_2=-1+2X$.
When $\xi_1>0$ and $\xi_2 \ge 0$, corresponding to $X \ge 1/2$,
the system is completely stable at the quantum level.
In the region of $0 \le X < 1/6$ one has $\xi_1<0$ and $\xi_2<0$ so that 
the perturbations are classically stable due to the 
positive sign of $c_s^2$. This vacua is, however, generally
quantum mechanically unstable.

The kind of instability prevented by the condition (\ref{xi2}), 
is of the tachyonic type and generally 
much less dramatic, provided that the conditions (\ref{xi}) are satisfied. 
The point is that if $p_{\vp\vp}>0$ there are modes, the ones with 
$k^2 < p_{\vp \vp}/p_X$, 
that undergo a classical exponential growth. This is an intrinsically IR
instability of the kind often encountered in cosmology when  
perturbations of super-horizon scales are considered. 
In fact, for a ``normal'' 
cosmological scalar field $p=X-V(\vp)$, one has just
$- p_{\vp \vp} = V_{\vp \vp} \simeq H_0^2$ where
$H_0$ is the Hubble rate.
For the above reasons we will adopt (\ref{xi}) but not (\ref{xi2}) as 
fundamental criteria 
for the consistency of the theory and we will eventually make sure 
that the effects of $\xi_3$ are negligible on the relevant physical scales.

\subsection{Scaling solutions} \label{sec_scaling}

We now set the field $\vp$ in a cosmological framework and study its 
evolution 
in a spatially flat Friedmann-Robertson-Walker (FRW) background spacetime 
with scale factor $a(t)$
\begin{equation}
\label{friedman}
ds^2=-dt^2+a^2(t)d{\bf x}^2\, .
\end{equation}
The equation for $\vp$ can be written  as
\begin{equation}
\label{geneeq}
{\ddot \vp}\left(p_X + \dot \vp ^2 p_{XX} \right) 
+ 3 H p_X \, \dot \vp
+ 2Xp_{X \vp} 
- p_\vp = -\sigma\,,
\end{equation}
where $H \equiv \dot{a}/a$ is the Hubble rate and 
the scalar charge $\sigma$ corresponds to the coupling 
between the matter and the dilaton
and it is defined by the relation 
$\delta {\cal S}_m/\delta \vp = - \sqrt{-g}\, \sigma$.
 
We are now interested in a late-time attractor of the kind studied in Refs.
\cite{wetter,Amendola2,GPV} i.e. characterized by a constant $\Omega_\vp$ 
and a constant (and, possibly, positive) acceleration. 
Thus we assume that the Universe is filled by the scalar field 
$\vp$ and by only one type of matter, of energy density  
$\rho_m$ and general equation of state $w_m$. In the following we
will also specialize the formula to the most relevant case $w_m = 0$, 
i.e. that of cold dark matter. 
By rewriting Eq.~(\ref{geneeq}) in terms of 
the energy density $\rho$ of the scalar field 
and then the equation for $\rho_m$ we obtain 
the system 
\begin{eqnarray}
\label{geneeq1}
\frac{\rd \rho}{\rd N}+3(1+w_\vp)\rho
=  - Q\rho_m \frac{\rd \vp}{\rd N}\,,\\[2mm]
\label{geneeq2}
 \frac{\rd \rho_m}{\rd N} + 3(1+w_m) \rho_m =  Q \rho_m 
 \frac{\rd \vp}{\rd N}\,,
\end{eqnarray}
where
\begin{equation}
\label{def}
N \equiv {\rm ln}\,a\,,~~~
w_{\vp} \equiv p/\rho\,,~~~
Q(\vp) \equiv \sigma /\rho_m\,.
\end{equation}

We also define in the usual way the fractional density of both components, 
$\Omega_m \equiv \rho_m/(3 H^2)$ and $\Omega_\vp \equiv \rho/(3 H^2)$, 
with $\Omega_\vp+ \Omega_m =1$.
We are looking for asymptotic scaling solutions where both the equation
of state parameter $w_\vp$ and the fractional density  $\Omega_\vp$
are constant during the evolution. 
In other words this corresponds to
${\rm d} \log \rho/{\rm d}N = {\rm d}\log \rho_m/{\rm d}N $, 
and by assuming that neither $Q$ is time-varying 
in the scaling regime we obtain, from (\ref{geneeq1}) 
and (\ref{geneeq2}), the relation
\begin{equation} 
\label{dphi}
\frac{\rd \vp}{\rd N} = \frac{3\Omega_\vp}{Q} 
(w_m - w_\vp) = {\rm const.}
\end{equation}

{}Inserting Eq.~(\ref{dphi}) back into Eqs.~(\ref{geneeq1}) 
and (\ref{geneeq2}),  
we get the scaling behavior of $\rho$ and $\rho_m$:  
\begin{equation} 
\label{sca}
\frac{\rd {\rm log} \rho}{\rd N}=\frac{\rd {\rm log} \rho_m}{\rd N}=
-3(1+w_s)\,,~~~{\rm with}~~~
w_s \equiv w_m+\Omega_\vp (w_\vp-w_m)\,.
\end{equation}
Here $w_s$ is the effective equation of state for each component 
and for the Universe itself along a scaling solution.
Note in fact that, because of the coupling terms in 
Eqs.~(\ref{geneeq1}) and (\ref{geneeq2}), $\rho$ and $\rho_m$ 
do not scale according to $w_\vp$ and $w_m$. 
By the definition of $X$ one also finds 
$ 2 X = H^2 (d\vp/dN)^2 \propto \, H^2$.
Therefore $X$ scales in the same way as $\rho$
and $\rho_m$, which means, by Eq.~(\ref{sca}), that 
\begin{equation} 
\label{X2eq}
\frac{\rd X}{\rd N}=-3(1+w_s) X \,.
\end{equation}
Of course, since the pressure density $p=w_\vp \rho$ has a same scaling 
behavior as $\rho$, one has $\rd p/\rd N=-3(1+w_s)$.
Since $p$ is also the Lagrangian of the scalar field and it is a
function of $X$ and $\vp$, 
this implies, by using Eqs.~(\ref{dphi}) and (\ref{X2eq}), that 
\begin{equation}
\label{pform}
\frac{\partial \log p}{\partial \log X} - 
\frac{1}{\lambda} \frac{\partial \log p}{d \vp} = 1\,,~~~
{\rm with}~~~\lambda\,  \equiv\, 
Q \frac{1+w_m - \Omega_\vp (w_m - w_\vp)}{\Omega_\vp
(w_m-w_\vp)}\,.
\end{equation}
The solution of this equation gives a constraint on the functional 
form of $p(X, \vp)$ for the existence of scaling solutions:
\begin{equation} 
\label{scap}
p(X, \vp) = X\,g\left(X e^{\lambda \vp}\right)\,,
\end{equation}
where $g$ is any function in terms of $ Y \equiv X e^{\lambda \vp}$.
By using Eqs.~(\ref{dphi}) and (\ref{X2eq}) one can easily 
show that $Y$ is constant along a scaling solution, say, $Y=Y_0$.

What we have found can be restated by saying that if the Lagrangian of a 
scalar field can be written in the form 
(\ref{scap}) and the scalar is coupled to matter with a coupling $Q$ which 
is weakly dependent of $\vp$, then a scaling solution may take place 
at late times. The ``geometrical'' properties of such a scaling solution 
are determined by the parameters $w_m$, $\lambda$ and $Q$ and are 
independent of the actual functional form of $g$. In fact, by the 
definition of $\lambda$ in Eq.~(\ref{pform}), the effective equation of state 
and the acceleration parameter $-q \equiv \ddot{a} a/{\dot a}^2$ are given 
by
\begin{equation} \label{w_s}
w_s = w_m - \frac{Q(1+w_m)}{\lambda + Q}, \qquad -q =  - \frac{3 w_s 
+1}{2} 
= \frac{3 Q (1+w_m) - (\lambda + Q)(3 w_m +1)}{2(\lambda +Q)} \,.
\end{equation}
Other quantities such as $w_\vp$ and $\Omega_\vp$, that also characterize 
the 
late time behavior of our model, depend on the actual form of the 
function $g$. As already noted, the 
argument of $g$ remains constant along the scaling solution, $Y = Y_0$.
A (last) relevant parameter to be introduced is therefore the log--log 
derivative of $g(Y)$ calculated on the attractor,
\begin{equation} \label{beta}
\beta \equiv \, \left. \frac{d \log g(Y)}{d \log Y} \right|_{Y = Y_0}.
\end{equation} 
Once the functional form of $g$ is given $\beta$ can be expressed as a 
function of the other parameters, as we will see in the following with some
example. The remaining quantities to be determined are then given by
\begin{equation} \label{womega}
w_\vp = \frac{1}{1+2\beta}\,, \qquad 
\Omega_\vp = \frac{Q(1+w_m)(1+2\beta)}{(\lambda + Q)[w_m(1+2\beta) - 1]} .
\end{equation}

\subsection{Scaling solutions with coupled dark matter 
(the case $w_m = 0$)}\label{dmatter}

Our main interest is the case in which the Universe is filled for the most
part with a non-relativistic (dark) matter ($w_m=0$) and the scalar field 
$\vp$. Since the dilaton follows the behaviour of the other component, in 
the 
limit $Q \to 0$ it behaves as a pressureless matter.
If we assume a non zero interaction between the dilaton and 
the matter ($Q \ne 0$), the scaling solutions can exhibit the 
acceleration of the Universe with nonzero $\Omega_\vp$.
It may be useful to specialize the relevant formulas of last subsection
to the $w_m=0$ case. Equation (\ref{w_s}) reduces to
\begin{equation} \label{w_s2}
w_s = w_\vp \Omega_\vp = - \frac{Q}{\lambda + Q}, \qquad -q 
= \frac{2Q -\lambda}{2(Q + \lambda)} .
\end{equation}
While the value of $\beta$ depends on the functional form of $g(Y)$,
we note that the actual numerical value $g_0$ of $g$ along the attractor is
constrained to be
\begin{equation} 
\label{g_0}
g_0 = - \frac{2}{3} Q(\lambda + Q) .
\end{equation}  
The above expression has been obtained by noting that 
$p=3 H^2 w_\vp \Omega_\vp$, by using the first equation in
(\ref{w_s2}) to express the product $w_\vp \Omega_\vp$ and, 
finally, by using the relation between $X$ and $H^2$: 
\begin{equation}
X \equiv {\dot \vp}^2 /2 = \frac{H^2}{2}\left(\frac{d\vp}{dN}\right)^2 = 
\frac{9 H^2}{2(\lambda + Q)^2} .
\end{equation}
The two remaining quantities are obtained directly from Eq.~(\ref{womega}):
\begin{equation} 
\label{w_vp2}
w_\vp = \frac{1}{1+2\beta}\,, \qquad 
\Omega_\vp = - \frac{Q(1+2\beta)}{\lambda + Q} .
\end{equation}
Since $\lambda$ is positive we get a positive acceleration for
$Q > \lambda/2$. The case $Q<-\lambda$ corresponds to 
a superinflationary attractor (with $-q >1$) which we will not consider 
here. 

\subsection{Simple cases}

It is useful to consider a couple of examples. 
Let us  start with a
Lagrangian of the form 
$p = f(X) - V(\vp)$. Then Eq.~(\ref{pform}) constrains $p$ to be
\begin{equation}
\label{simplesca}
p(X, \vp) = \epsilon X - c_2 e^{-\lambda \vp}\,,
\end{equation}
where the constant $\epsilon$ is either positive or negative.
This corresponds to choosing
the arbitrary function $g(Y)$ as $g(Y) = \epsilon - c_2/Y$. 
With a field redefinition we can always 
set $\epsilon=\pm 1$ without loss of generality. 
The scaling solutions for $\epsilon=1$ (normal scalar field)
were already extensively studied in the literature 
(e.g. \cite{Amendola2,GPV}), while a negative $\epsilon$ corresponds to 
the (quantum mechanically unstable) phantom scalar field with an 
exponential potential. 
The effective equation of state and acceleration of the 
Universe along the attractor are already given by eq. (\ref{w_s2}) once $Q$ and the 
slope of the potential $\lambda$ are given. In order to obtain
$\Omega_\vp$ and $\omega_\vp$ from (\ref{w_vp2}) we need an expression for 
the parameter $\beta$. 
By Eq.~(\ref{beta}) we have $\beta = (c_2/Y_0)/(\epsilon - c_2/Y_0)$. 
The value $Y_0$ is
related to $g_0$ by $c_2/Y_0 = \epsilon - g_0$ and $g_0$ is constrained 
by Eq.~(\ref{g_0}). 
We finally obtain
\begin{equation} 
\label{likebefore}
\beta = - \frac{3 \epsilon + 2 Q(\lambda + Q)}{ 2 Q(\lambda + Q)}\,, \quad
w_\vp= - \frac{Q(\lambda+Q)}{Q^2+Q\lambda+3\epsilon}\,, \quad
\Omega_\vp=\frac{Q^2+Q\lambda+3\epsilon}{(\lambda+Q)^2}\,.
\end{equation}
When $\epsilon = 1$ these formulas are in agreement with those found in 
Refs.~\cite{Amendola2,GPV}. 
Moreover, in the non-accelerating case with $Q=0$, one has 
$w_\vp=0$ and $\Omega_\vp=3/\lambda^2$, as found e.g. in \cite{Cope}. 
For the phantom type scalar field ($\epsilon=-1$), the fractional
energy density $\Omega_\vp$ is negative for $Q=0$ 
($\Omega_\vp=-3/\lambda^2$) which, by itself,
means that this case can hardly be considered as realistic.
This situation may be improved by implementing the 
coupling $Q$, although we need to caution that the system 
(\ref{simplesca}) with $\epsilon = -1$ is unstable at the quantum 
level unless higher-order terms in $X$ are taken into 
account. 

As a second example we consider a higher-order kinetic correction 
to the usual kinetic term,
\begin{equation}
p(X,\varphi) = c_1 X^2 e^{\lambda \varphi} - X,
\end{equation}
i.e. $g(Y) = c_1 Y -1$. By proceeding as before we obtain
\begin{equation} 
\beta = - \frac{3 - 2 Q(\lambda + Q)}{2 Q(\lambda + Q)}\,, \quad
w_\vp=\frac{Q(\lambda+Q)}{3 Q(\lambda + Q)-3}\,, \quad
\Omega_\vp=3 \frac{1 - Q(\lambda+Q)}{(\lambda+Q)^2}\,.
\end{equation}
In this case, the stability region is characterized by 
$2 c_1 X e^{\lambda \vp}>1$ from Eq.~(\ref{xi})  
and we have a stable late time attractor 
for $\lambda > Q -3/(4 Q)$.

\subsection{A kind of gauge symmetry}
We conclude this section by making a few comments on a ``gauge symmetry'' 
which 
underlies our equations. The theory of a scalar field with a Lagrangian 
$p(X,\vp)$
as in (\ref{lag2}) is, in fact, invariant under a general redefinition of 
the field
itself. This formally means that all the equations have to be invariant 
under
the transformation 
\begin{equation} 
\label{transf}
\begin{cases}
\vp \, \rightarrow \, {\widetilde \vp}(\vp) \ &= {\widetilde \vp}(\vp),\\ 
X \, \rightarrow \, \widetilde X(X,\vp) &= {\widetilde \vp}'(\vp)^2 X .
\end{cases}
\end{equation}
Equivalently, there are different Lagrangians that give the same physics.
Take for instance $p(X,\vp)$ and $\wt p(X,\vp)$ defined as 
\begin{equation}
\wt p(X,\vp) = p\left(\wt X(X,\vp), \wt \vp(\vp)\right) .
\end{equation}
They must be equivalent since $\wt p$ is obtained from $p$ by a simple 
redefinition of the field. 
All the evolution equations are in fact invariant under the 
transformation (\ref{transf}). 
Among the positive definiteness conditions for the Hamiltonian 
in Eqs.~(\ref{xi}) and (\ref{xi2}), 
only (\ref{xi}) are invariant under (\ref{transf}), 
while (\ref{xi2}) is not.
This reinforces the choice of not considering (\ref{xi2}) as a basic 
requirement
for the quantum stability of the system. 

Under the field redefinition (\ref{transf})
the scalar charge $Q$ defined in (\ref{def}) 
varies according to 
\begin{equation} 
Q(\vp) = Q(\wt \vp)\frac{{\rm d} \wt \vp}
{{\rm d} \vp} .
\end{equation}
Therefore, when considering scaling solutions in section 
\ref{sec_scaling}, 
we ``fixed'' the
gauge freedom by taking the charge $Q$  as a constant function of $\vp$. 
In other words, taking $Q$ independent of $\vp$
is not a restriction, but rather,  
identifies the scalar field $\vp$ that we are speaking of. 
The functional form of the Lagrangian (\ref{scap}) has thus to be intended 
as 
referred to that particular ``gauge fixed'' scalar field.
Of course, there is a residual gauge freedom: a rescaling of the type 
$\vp \rightarrow \alpha \, \vp$, with $\alpha$ a positive number. 
The parameters of the scaling solutions 
change, accordingly, as $Q\rightarrow Q/\alpha,\ \lambda\rightarrow 
\lambda/\alpha, \
\beta \rightarrow \beta$ and formulas (\ref{w_s}) and 
(\ref{womega}) are invariant under such a rescaling.

\section{A phase space analysis of various cosmological scenarios}

In this section we shall study the cosmological evolution 
for the string-inspired Lagrangian (\ref{ourlag}).
The dynamics of our cosmological ghost--dilaton model is best 
studied by rewriting the equations as an autonomous system. 
We note that, in the special case where $\lambda_1 = \lambda$, 
the Lagrangian (\ref{ourlag}) is included as one of the scaling 
solutions with the choice
\begin{equation} 
\label{gYphan}
g(Y)=-1+c_1Y-c_2/Y\,.
\end{equation}
The basic equations we need are the Friedmann equation
\begin{equation}
\label{backeq}
3H^2=-\frac12 \dot{\vp}^2+\frac34c_1 e^{\lambda \vp}
\dot{\vp}^4+c_2e^{-\lambda \vp}+\rho_m\,,
\end{equation} 
and its first derivative
\begin{equation}
\label{backeq2}
 2\dot{H}=\dot{\vp}^2-c_1e^{\lambda \vp}
\dot{\vp}^4-\rho_m\,.
\end{equation}
The latter is equivalent to the evolution equation (\ref{geneeq}).

In order to 
write the above equations with an autonomous system
it is convenient to introduce the 
following quantities:
\begin{equation} 
\label{quan}
x^2 \equiv \frac{\dot{\vp}^2}{6H^2}\,,~~~
z^2 \equiv \frac{e^{-\lambda \vp}}{3H^2}\,,
\end{equation}
by which Eq.~(\ref{backeq}) can be rewritten as
\begin{equation} 
\label{rehubble}
\Omega_\vp+\Omega_m=1\,,~~~
{\rm with}~~~\Omega_\vp=-x^2+3c_1x^4/z^2+c_2z^2 \, .
\end{equation}
Following Ref.\,\cite{Cope} we shall rewrite the evolution
equations in terms of $x$
and $z$. 
We assume that the matter $\rho_m$ is
non-relativistic ($p=0$).
Then from Eqs.~(\ref{backeq}) and (\ref{backeq2}) we obtain 
\begin{equation} 
\label{dehubble}
\frac{\rd H}{\rd N}=\frac32H
\left(x^2-c_1 \frac{x^4}{z^2}+c_2z^2-1\right)\,.
\end{equation}
Making use of Eqs.~(\ref{geneeq}) and (\ref{dehubble}), 
we find
\begin{eqnarray}
\label{xeq}
\frac{\rd x}{\rd N}&=&-\frac32 \left(x^2-
c_1\frac{x^4}{z^2}+c_2z^2-1\right)x
+\frac{\sqrt{6}z^2}{2(6c_1x^2-z^2)} \nonumber \\
&\times&
\left[-Q \left(1+x^2-3c_1\frac{x^4}{z^2}
-c_2z^2\right)-\sqrt{6}x\left(c_1 \frac{2x^2}{z^2}
-1\right)-\lambda c_1 \frac{3x^4}{z^2}
+\lambda c_2 z^2 \right], \\
\label{zeq}
\frac{\rd z}{\rd N}&=& -\frac{\sqrt{6}}{2}
\lambda xz-\frac32 z \left(x^2-c_1\frac{x^4}{z^2}
+c_2z^2-1\right)\,.
\end{eqnarray}
Once the initial conditions of $x$ and $z$ are known, 
we get the values of $\Omega_\vp$ and $\Omega_m$
by solving Eqs.~(\ref{xeq}) and (\ref{zeq}).

Basically our main interest is the case $c_1 \ne 0$ and $c_2=0$,
since the stability condition (\ref{xi}) is satisfied even for $c_2=0$.
This is a minimal string-inspired scenario which ensures 
the stability of the system.
In order for the generality of our discussion, however, we also 
implement the $c_2$ term corresponding to the exponential 
potential of the dilaton. If we impose the stability condition 
(\ref{xi2}) in addition to (\ref{xi}), it is required to take 
into account the nonvanishing potential ($c_2 \ne 0$).
Although the condition (\ref{xi2}) is not obligatory compared to 
(\ref{xi}) in order to ensure the stability of quantum fluctuations, 
we shall discuss the effect of the exponential potential 
in the later subsection.
Hereafter we consider the case $\lambda=\lambda_1$.

In subsection \ref{config} we analyze the parameter range 
in terms of $(x^2, z^2)$ in which the stability of quantum 
fluctuations is satisfied. In subsection \ref{viable} it is 
shown that a viable cosmological scenario can be obtained for $Q=0$
without violating the stability of the system.
This is one of the most important results of our work. 
In subsection \ref{nonzeroQ} we shall study the case with 
nonzero $Q$ and show that it is difficult to get an appropriate
cosmological evolution unless $Q$ is a rapidly growing function.
In Appendix we shall discuss the classical dynamics of the 
phantom field in the absence of the higher-order kinematic term 
($c_1=0$).

\subsection{Restrictions on the configuration space} \label{config}

In this subsection we derive the conditions that our new variables 
need to satisfy in order to ensure the stability of the system.
{}From the condition (\ref{xi}) we obtain 
\begin{equation} 
\label{stabi}
z^2 \le 2c_1x^2\,,
\end{equation}
where Eq.~(\ref{Ygeneral}) is used.
Note that this is also written as $Y=x^2/z^2 \ge 1/(2c_1)$.
In what follows we shall consider two cases (A) $c_2=0$
and (B) $c_2 \ne 0$ separately.

\subsubsection{Case of $c_2=0$}

When the potential of the dilaton is absent ($c_2=0$), 
the requirement of the condition,
$\Omega_\vp=-x^2+3c_1x^4/z^2\le 1$, yields
\begin{equation} 
\label{requ1}
z^2 \ge \frac{3c_1x^4}{x^2+1}\,.
\end{equation}
Note that $\Omega_\vp \ge 0$ is automatically
satisfied under the condition of Eq.~(\ref{stabi}).
In Fig.~\ref{stability} we plot the region characterized
by Eqs.~(\ref{stabi}) and (\ref{requ1})
in the $(x^2, z^2)$ plane.
The two parameters $x^2$ and $z^2$ need be 
inside this region in order to 
ensure the stability of system.
Note that in the case of $c_2=0$ the condition (\ref{xi2})
is not satsified. However this condition is not fundamentally 
required for the stability of quantum fluctuations.

\subsubsection{Case of $c_2 \ne 0$}

\begin{figure}
\epsfxsize = 6.0in \epsffile{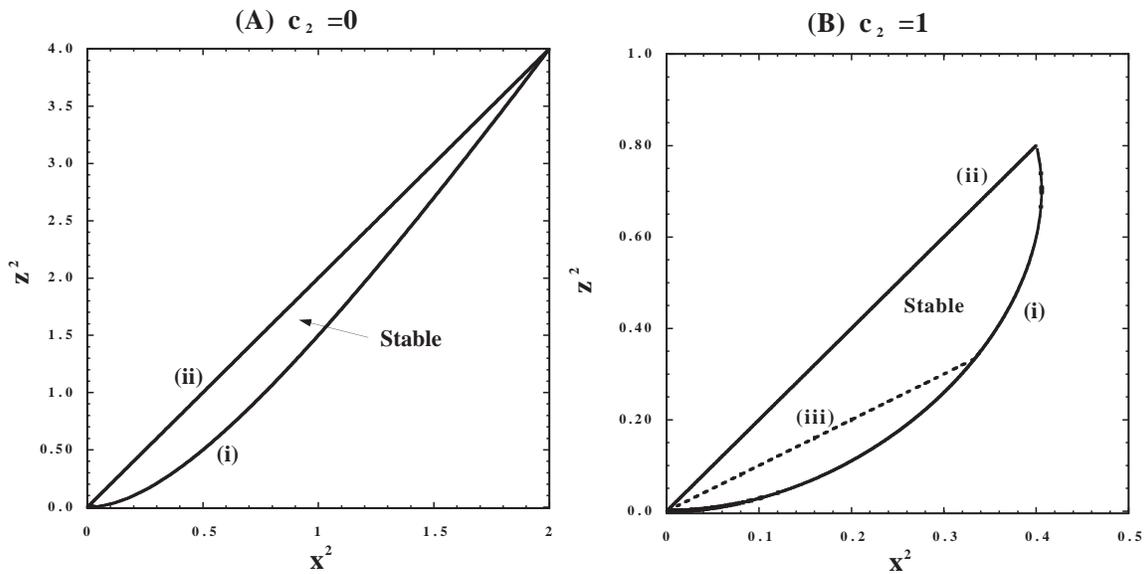} 
\caption{The region in which the conditions
(\ref{xi}) and $\Omega_\vp \le 1$ are satisfied
for (A) $c_1=1$, $c_2=0$ and (B) $c_1=1$, $c_2=1$.
In the case (A) each curve or line corresponds to 
(i) $z^2=(3c_1x^4)/(x^2+1)$ and 
(ii) $z^2=2c_1x^2$, whereas  in the case (B)
each corresponds to (i) $z^2=\frac{1}{2c_2} 
\left[x^2+1 \pm \sqrt{(x^2+1)^2-12c_2x^4}
\right]$, (ii) $z^2=2c_1x^2$
and (iii) $z^2=\sqrt{c_1/c_2}\,x^2$.
}
\label{stability}
\end{figure}

The constrained region changes in the presence of 
the dilaton potential ($c_2 \ne 0$).
In addition to the condition (\ref{stabi}) we obtain 
the following relation from the requirement $\Omega_\vp \le 1$:
\begin{equation} 
\label{zcon}
\frac{1}{2c_2} \left[x^2+1-\sqrt{(x^2+1)^2-12c_2x^4}
\right] \le z^2 \le \frac{1}{2c_2} \left[x^2+1+
\sqrt{(x^2+1)^2-12c_1c_2x^4}\right]\,.
\end{equation}
The stability region (\ref{stabi}) is plotted 
together with (\ref{zcon}) in the right panel of 
Fig.~\ref{stability} for $c_2=1$.
Unlike the case of $p=X-c_2e^{-\lambda \vp}$
in which the constrained regime is 
$x^2+z^2 \le 1$ \cite{Cope}, 
the allowed phase space is more restricted.
Note that there exist allowed regions even in 
the presence of the condition (\ref{xi2}), i.e.
$\sqrt{c_1/c_2}\,x^2 \le z^2$.
This happens for nonzero values of $c_2$, but 
Eq.~(\ref{xi2}) is not obligatory as we already 
mentioned.

%
\subsubsection{Equation of state $w_\vp$}

We wish to point out general features when we impose 
the stability condition (\ref{xi}).
The $p_X+\dot{\vp}^2p_{XX}$
term in Eq.~(\ref{geneeq}) is positive as long as 
Eq.~(\ref{xi}) holds.
This means that the dilaton does not climb up the potential hill
unlike the case of the classical evolution discussed in the Appendix.
The equation of state for the field $\vp$ is given as 
\begin{equation} 
\label{wp}
w_\vp=\left(2\frac{X}{p}\frac{\partial p}{\partial X}-1
\right)^{-1}\,.
\end{equation}
In order to achieve the equation of state with 
$w_\vp=p/\rho<0$, we require the negative 
pressure ($p<0$). If we impose the condition 
(\ref{xi}), one has 
$2(X/p)(\partial p/\partial X) \le 0$ and 
$w_\vp \ge -1$.
Therefore as long as the stability condition
(\ref{xi}) is imposed, one gets the equation of 
state for the normal scalar field,  
in spite  of the fact that 
the coefficient of the $\dot{\vp}^2$ term is negative.
This is a general feature of any form of the Lagrangian $p$.

\subsection{Dynamics of the system for $c_1 \ne 0$ and $Q=0$} \label{viable}

Let us consider the dynamical evolution for the system 
characterized by the Lagrangian (\ref{ourlag}) with $Q=0$.
In the next subsection we shall study the case in which the 
coupling $Q$ is taken into account.

The discussion in Sec.\,\ref{sec_scaling} demonstrates that one has a
scaling solution where $p$ is vanishing for $Q=0$.
This is not an interesting attractor solution, 
since it does not lead to an acceleration of the Universe.
As studied in Appendix, it is expected that there exist
other attractor solutions where $\Omega_\vp$ approaches 1. 
These solutions can be found by setting $\rd x/\rd N=0$
and $\rd y/\rd N=0$ in Eqs.~(\ref{xeq}) and (\ref{zeq}).
Especially we are interested in the case where $w_\vp$
approaches a negative constant close to $-1$.
{}From Eq.~(\ref{wp}) the equation of state for the 
Lagrangian (\ref{ourlag}) is given by 
\begin{equation} 
\label{wgene}
w_\vp=\frac{-1+c_1Y-c_2/Y}{-1+3c_1Y+c_2/Y}\,.
\end{equation}
Then one may look for the attractor solutions which keep
the quantity $Y=x^2/z^2$ constant.
Setting $\rd x/\rd N=0$ and $\rd y/\rd N=0$
in Eqs.~(\ref{xeq}) and (\ref{zeq}), we get 
\begin{equation} 
\label{xandxs}
x=\frac{\sqrt{6}(2c_1Y-1)}{\lambda(3c_1Y-1+c_2/Y)}\,,~~
{\rm and}~~
x^2=\frac{1-\sqrt{6}\lambda x/3}{1-c_1Y+c_2/Y}\,.
\end{equation}
Eliminating $x$ from these equations gives
\begin{eqnarray} 
\label{Yeq}
& & 24c_1^3 Y^5-3c_1^2(16+\lambda^2)Y^4
-(24c_1^2c_2-30c_1-4\lambda^2c_1)Y^3 \nonumber \\
& &+(24c_1c_2-6-\lambda^2+2\lambda^2c_1c_2)Y^2
-6c_2Y+\lambda^2c_2^2=0\,.
\end{eqnarray}

In the limit of $\lambda \to 0$ one obtains the following solutions
\begin{equation} 
\label{solu}
Y=0,~\frac{1}{2c_1},~\frac{1 + \sqrt{1+4c_1c_2}}{2c_1}\,,
\end{equation}
where we used $Y=Xe^{\lambda \vp} \ge 0$.
The ghost condensate considered in Ref.~\cite{Arkani}
corresponds to the attractor $Y=1/(2c_1)$ 
with $c_2=0$.
Generally we have two stable 
solutions, i.e. $Y=1/(2c_1)$ and $Y=(1+\sqrt{1+4c_1c_2})/(2c_1)$.
The former one is an attractor solution in which 
the system evolves toward the de-sitter phase characterized
by a cosmological constant ($w_\vp=-1$).
The latter one is the scaling solution for $Q=0$
discussed in Sec.~\ref{sec_scaling}, but this does not correspond to
an accelerating universe since the equation of state is 
$w_\vp=0$.

In the case of nonzero $\lambda$, it is generally not easy to 
get an explicit form of solutions for Eq.~(\ref{Yeq}).  
However we can obtain numerical values of $Y$ once
$c_1$, $c_2$ and $\lambda$ are specified in Eq.~(\ref{Yeq}).  
In what follows we shall consider the cases (A) $c_2=0$
and (B) $c_2 \ne 0$ separately.

\subsubsection{Case of $c_2=0$}

When $c_2=0$ Eq.~(\ref{Yeq}) gives
\begin{equation} 
\label{solu2}
Y=0,~~\frac{1}{c_1},~~
\frac{1}{c_1}\left[\frac12+\frac{\lambda^2}{16}
\left(1 +\sqrt{1+\frac{16}{3\lambda^2}}\right)
\right],~~\frac{1}{c_1}\left[\frac12+\frac{\lambda^2}{16}
\left(1 -\sqrt{1+\frac{16}{3\lambda^2}}\right)
\right]\,.
\end{equation}
Among them the solutions satisfying the condition  
(\ref{xi}) are the second and third ones.
The former one is a non-accelerating scaling solution
with $w_\vp=0$. 
In the latter case the equation of state (\ref{wgene}) is written 
in terms of $\lambda$:
\begin{equation} 
\label{eqlam}
w_\vp=-\frac{1-(\lambda^2/8)-(\lambda/24)\sqrt{9\lambda^2+48}}
{1+(3\lambda^2/8)+(\lambda/8)\sqrt{9\lambda^2+48}}\,.
\end{equation}
In the limit of $\lambda \to 0$ we have $w_\vp \to -1$.
The equation of state ranges in the acceleration region $-1 \le w_\vp \le -1/3$
as long as $\lambda~\lsim~0.82$.
For example one has $w_\vp=-0.889$ for $\lambda=0.1$.

\begin{figure}
\epsfxsize = 4.0in \epsffile{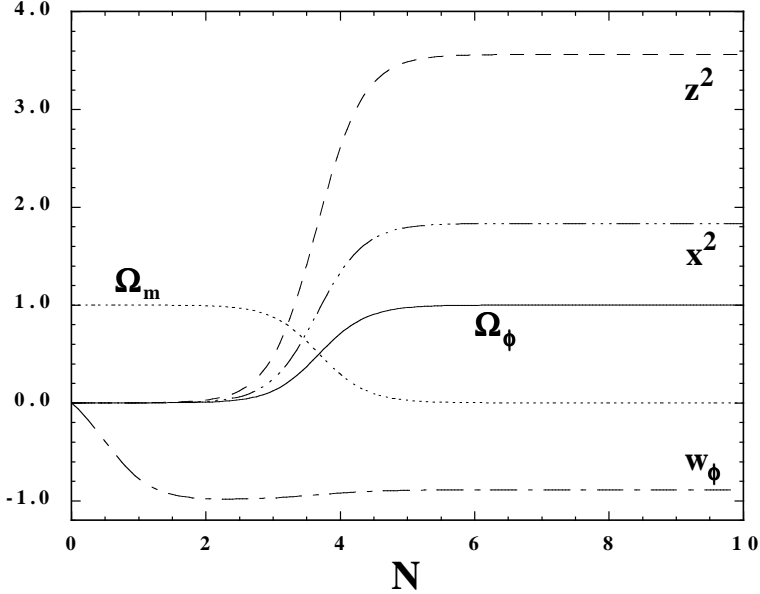}
\caption{The variation of $\Omega_m$, $\Omega_\vp$, $w_\vp$, $x^2$ and
$z^2$ for $c_1=1$, $c_2=0$ and $\lambda=0.1$
with initial conditions $x_i = 0.0085$ and $z_i =0.0085$.
The solution evolves toward
the attractor characterized by constant $x^2$ and $z^2$
with $\Omega_\vp \to 1$ and $\Omega_m \to 0$. 
The equation of state asymptotically approaches
a constant value $w_\vp \simeq -0.889$.
}
\label{evon}
\end{figure}

In Fig.~\ref{evon} we plot the evolution of 
$\Omega_m$, $\Omega_\vp$, $w_\vp$, $x^2$ and
$z^2$ for $c_1=1$, $c_2=0$ and $\lambda=0.1$.
The solution evolves toward the attractor characterized by constant 
$x^2$ and $z^2$ with $\Omega_\vp \to 1$ and $\Omega_m \to 0$.
The third solution in Eq.~(\ref{solu2}) gives $Y \simeq 0.515$
for $c_1=1$ and $\lambda=0.1$.
Making use of Eq.~(\ref{xandxs}) and $Y=x^2/z^2 \simeq 0.515$, 
we obtain $x^2 \simeq 1.834$ and $z^2=3.561$, in which case
$\Omega_\vp=-x^2+3c_1x^4/z^2$ asymptotically approaches 1.
This is different from the scaling solution that approaches
a nonzero value of $\Omega_\vp$ or $\Omega_m$ discussed
in Sec.\,\ref{sec_scaling}. 
Note that the discussion in Sec.\,\ref{sec_scaling} does not apply to 
the case in which $\Omega_m$ or $\Omega_\vp$ approaches zero 
asymptotically.

Numerical integrations of 
Eqs.~(\ref{xeq}) and (\ref{zeq}) agree very well with our analytic 
estimates, see Fig.~\ref{evon}. 
We begin to integrate the equations with initial conditions
satisfying $x^2, z^2 \ll 1$ from the epoch where matter 
dominates and radiation can be neglected.
The case shown in Fig.~\ref{evon} corresponds to the initial matter 
density $\rho_m \simeq 9.0 \times 10^{-119}M_p^4$, 
in which case $\Omega_\vp$ grows to 0.7 at the present 
epoch $N \simeq 4$.
The system asymptotically evolves toward the 
values $\Omega_\vp=1$, $\Omega_m=0$, $x^2=1.834$ and $z^2=3.561$.
We found that the results hardly change even if the radiation is 
included at the beginning.
We also checked that 
the system always lies inside the quantum mechanically allowed
region shown in the left panel of Fig.~\ref{stability}.
This is also confirmed in Fig.\,\ref{evon} in which the equation of 
state is larger than $-1$.

\begin{figure}
\epsfxsize = 4.0in \epsffile{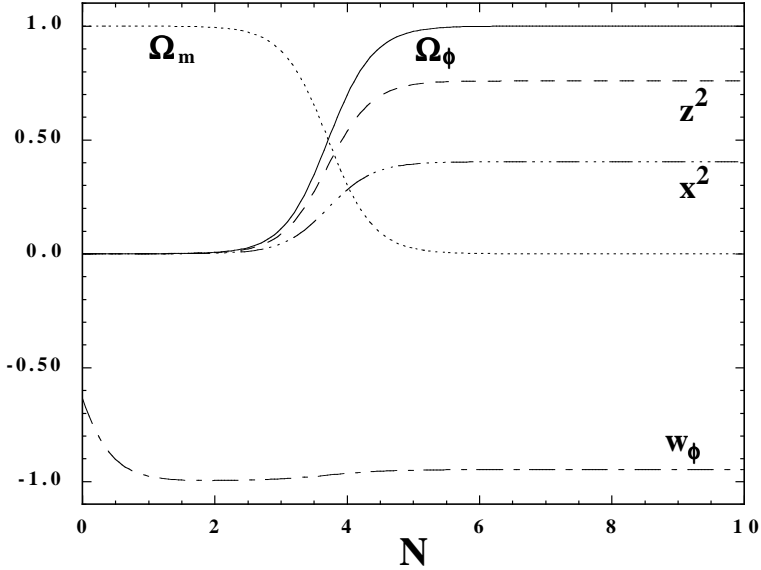}
\caption{The variation of $\Omega_m$, $\Omega_\vp$, $w_\vp$, $x^2$ and
$z^2$ for $c_1=1$, $c_2=1$ and $\lambda=0.1$
with initial conditions $x_i = 0.003$ and $z_i =0.0035$.
The evolution of $\Omega_\vp$ and $\Omega_m$ is similar 
to Fig.~\ref{evon}, although the asymptotic values of 
$x^2$, $z^2$ and $w_\vp $ are different.}
\label{evo}
\end{figure}

%
\subsubsection{Case of $c_2 \ne 0$}

Let us next consider the situation where the exponential potential of the dilaton 
is present ($c_2 \ne 0$).
Equation (\ref{Yeq}) is more involved than in the case of $c_2=0$, but
we can numerically solve Eq.~(\ref{Yeq})
for given values of $\lambda$, $c_1$ and $c_2$.
For example, when $\lambda=0.1$ and $c_1=c_2=1$,
the solutions are $Y\simeq 0.00167, 0.467, 0.532, 1.62$.
Since the last one ($Y \simeq 1.62$) corresponds to the non-accelerating 
universe, only the third one ($Y \simeq 0.532$) is an important solution 
satisfying the stability condition (\ref{xi}).
In this case the equation of state is given as
$w_\vp \simeq -0.948$ by Eq.~(\ref{wgene}), 
which is slightly larger than $-1$.
We also obtain the asymptotic values 
$x^2=0.4039$ and $z^2=0.7591$, in which case the 
fractions of
the kinematic energy and the potential energy of the dilaton  
are $\Omega_K=-x^2+3c_1x^4/z^2 \simeq 0.2409$ and 
$\Omega_V=c_2z^2 \simeq 0.7591$, respectively.
Therefore $\Omega_\vp=\Omega_K+\Omega_V$ asymptotically 
approaches unity. The numerical simulation in Fig.~\ref{evo}
agrees well with these asymptotic values of 
$w_\vp$, $x^2$, $z^2$ and $\Omega_\vp$.

By taking into account the nonzero $c_2$ term, the allowed 
region in the $(x^2, z^2)$ plane tends to shift toward smaller 
values of $x^2$ and $z^2$ as seen in Fig.~\ref{stability}. 
This does not necessarily mean that the cosmological evolution 
is more restricted than in the case of $c_2=0$.
In the case shown in Fig.~\ref{evo}
the system always lies inside the quantum mechanically allowed
region shown in Fig.~\ref{stability}.
In addition even the condition (\ref{xi2}) is satisfied 
throughout the evolution.

In summary we can obtain stable cosmological  
solutions which asymptotically 
approach $\Omega_\vp=1$ and $\Omega_m=0$ in the future
both in the case of $c_2=0$ and $c_2 \ne 0$.
We stress that this is different from the scaling solutions
discussed in Sec.\,3.

\subsection{Dynamics of the system for $c_1 \ne 0$ and $Q \ne 0$} 
\label{nonzeroQ}

If there is a coupling between the dilaton and the matter ($Q \ne 0$), 
one has additional scaling solutions discussed in Sec.~\ref{sec_scaling}.
When $w_m=0$ we have $w_\vp =-Q/(\Omega_\vp (\lambda+Q))$
from Eq.~(\ref{w_s2}) in the scaling regime.

Using the relations $w_\vp=p/\rho$, $\Omega_\vp=\rho_\vp/(3H^2)$,
$3H^2=2(\lambda+Q)^2X/3$ and $p=X(-1+c_1Y-c_2/Y)$, 
Eq.~(\ref{w_s2}) gives the constant value of $Y$:
\begin{equation} 
\label{YQ}
Y=\frac{1}{2c_1} \left[1-\frac23 Q(\lambda+Q)+
\sqrt{\left[1-\frac23 Q(\lambda+Q)\right]^2+4c_1c_2}\right]\,.
\end{equation}
Note that the third solution in Eq.~(\ref{solu}) is recovered in the limit 
of $Q \to 0$.
Substituting Eq.~(\ref{YQ}) for $\Omega_\vp=\rho_\vp/(3H^2)$
with $\rho_\vp=X(-1+3c_1Y+c_2/Y)$, we obtain
\begin{equation} 
\label{Omephige}
\Omega_\vp=\frac{3}{(\lambda+Q)^2}\left[1-Q(\lambda+Q)
+\frac{12c_1c_2}{3-2Q(\lambda+Q)+\sqrt{[3-2Q(\lambda+Q)]^2
+36c_1c_2}}\right]\,.
\end{equation}
The equation of state $w_\vp$ is determined from Eqs.~(\ref{w_s2})
and (\ref{Omephige}).

If one requires the late time behavior with $\Omega_\vp \simeq 0.7$
and $w_\vp \simeq -1$, Eq.~(\ref{w_s2}) gives the relation 
$\lambda=(3/7)Q$.
Substituting this relation for Eq.~(\ref{Omephige}), 
we can determine the values of $Q$ and $\lambda$
for known values of $c_1$ and $c_2$.
For example, in the case of $c_1=1$ and $c_2=0$, 
one has $Q=0.678$ and $\lambda=0.398$ for 
$\Omega_\vp \simeq 0.7$ and $w_\vp \simeq -0.9$.
As another example, when $c_1=1$ and $c_2=1$,  
$Q \simeq 1.45$ and $\lambda \simeq 0.85$
for $\Omega_\vp \simeq 0.7$ and $w_\vp \simeq -0.9$.

\subsubsection{Case of constant $Q$}

\begin{figure}
\epsfxsize = 3.5in \epsffile{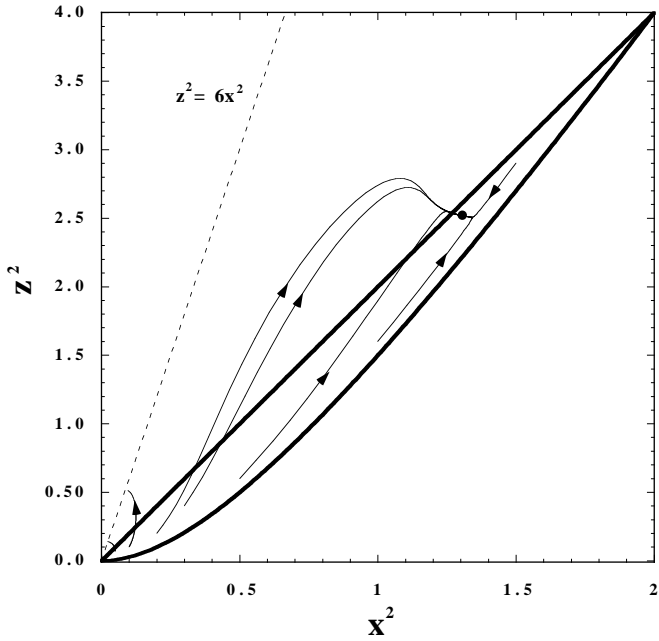} 
\caption{The phase-space trajectories for
$Q=0.678$ and $\lambda=0.398$ for $c_1=1$ and $c_2=0$.
In this case there exist scaling solutions that asymptotically approach
$\Omega_\vp \simeq 0.7$ and $w_\vp \simeq 
-0.9$, as long as the initial values of  $x^2$ and $z^2$ are not far from 
the attractor point $(x_s^2, z_s^2)=(1.295, 2.522)$ (this point is indicated 
by a black point in the figure).
If the initial values of  $x^2$ and $z^2$ are much smaller than 1, 
the trajectories tend to evolve out of the stable region.
In particular it happens that the solutions hit the singularities with $z^2=6x^2$
at which the speed of sound diverges.
}
\label{pspace}
\end{figure}

\begin{figure}
\epsfxsize = 3.5in \epsffile{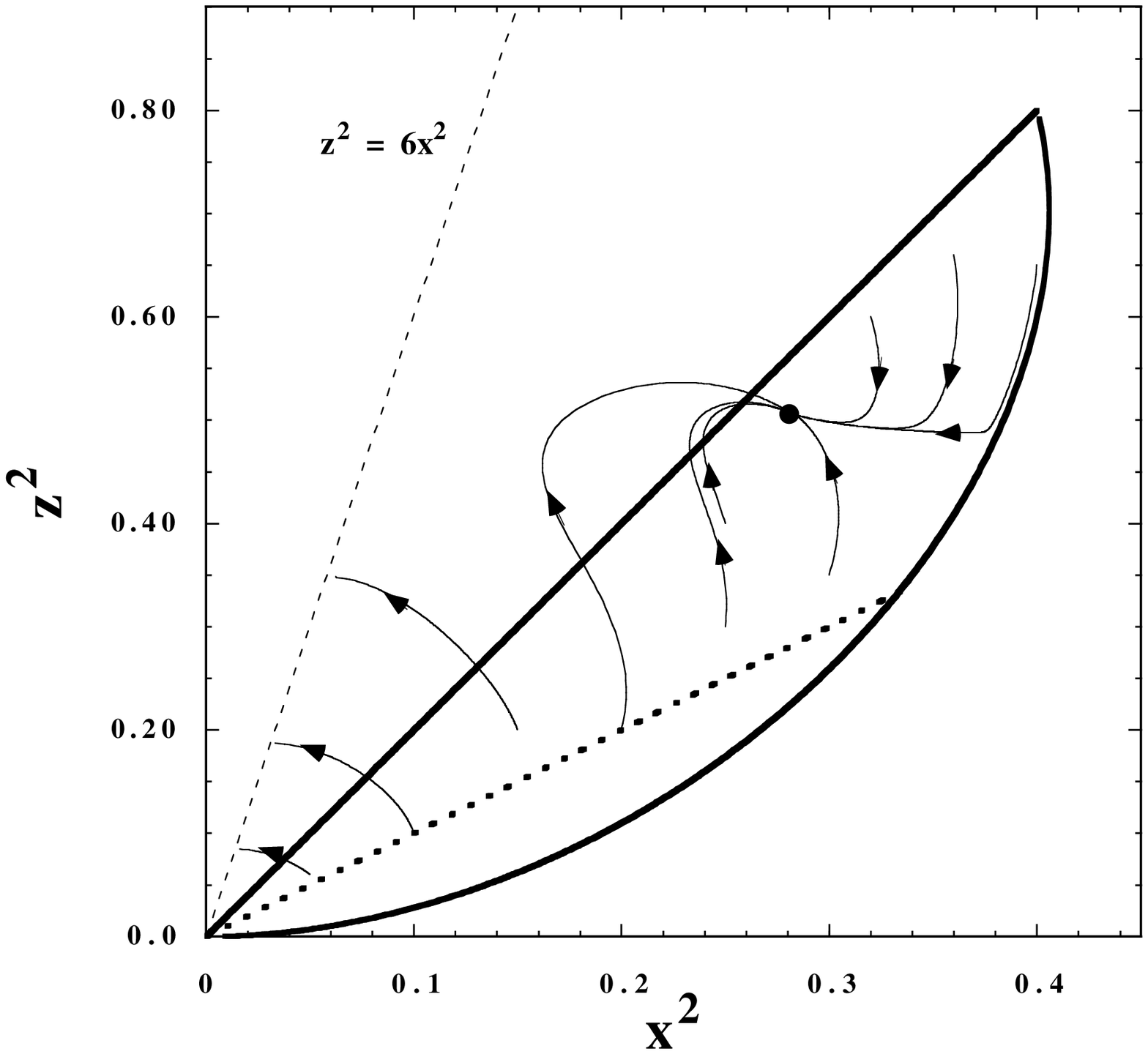} 
\caption{The phase-space trajectories for 
$Q=1.45$ and $\lambda=0.85$ for $c_1=1$ and $c_2=1$.
In this case scaling solutions correspond to 
$x_s^2=0.283$ and $z_s^2=0.505$ with  asymptotic values 
$\Omega_\vp \simeq 0.7$ and $w_\vp \simeq -0.9$.
As is similar to Fig.~\ref{pspace}
the trajectories tend to be away from the stable region 
when the initial values of  $x^2$ and $z^2$ are much smaller than 1.
}
\label{phase}
\end{figure}

Let us consider the situation in which the coupling $Q$ is constant 
not only in the scaling regime but also in the matter dominated era.
We first analyze the case where the exponential 
potential is absent ($c_2=0$).
In Fig.~\ref{pspace} we plot the phase-space trajectories for 
$Q=0.678$, $\lambda=0.398$ and $c_1=1$
with different initial conditions of $x^2$ and $z^2$. 
One has an attractor
point $(x_s^2, z_s^2)=(1.295, 2.522)$ corresponding to the scaling 
solution given in Eqs.~(\ref{w_s2}) and (\ref{Omephige}).
In order to ensure the stability of quantum fluctuations, the trajectories
are required to be inside the region shown in Fig.~\ref{stability}.
As seen from Fig.~\ref{phase}, only the initial conditions which are not 
far from the attractor point $(x_s^2, z_s^2)$ satisfy this 
constraint.
There are some trajectories which finally approach the attractor 
point but are away from the stability region in the middle of the way.
In this case the vacuum is unstable at the quantum level 
during some moment of time, 
so we can not regard this as ideal scaling solutions.

During the matter dominant phase $x^2$ and $z^2$ are needed to be much 
smaller than unity. This means that the trajectories start from the region with 
$x^2, z^2 \ll 1$ as a realistic dark energy model.
However, these trajectories do not satisfy the stability condition 
as seen in Fig.~\ref{pspace}. 
In addition they fall into the singularity on the line $z^2=6c_1x^2$ at which 
the speed of sound diverges.
It is inevitable to avoid this divergent behavior as long as $x^2$ is 
initially less than of order $0.1$. 

This situation does not change much even when the exponential potential of 
the dilaton is present ($c_2 \ne 0$).
In Fig.~\ref{phase} the phase-space trajectories are shown for 
$Q=0.678$, $\lambda=0.398$, $c_1=1$ and $c_2=1$.
In this case the attractor point corresponds to 
$(x_s^2, z_s^2)=(0.283, 0.505)$, whose values are smaller
than in the case of $c_2=0$.
However the behavior of solutions with initial conditions
$x_i^2, z_i^2 \ll 1$ is similar to the one discussed above,
which means that the trajectories hit the singularities at 
$z^2=6c_1x^2$ in addition to the violation of 
the stability condition.
Therefore  inclusion of the {\it constant} coupling $Q$ does
not help to obtain realistic scaling solutions where $\Omega_\vp$ 
is initially much smaller than 1 and evolves to a constant value 
$\Omega_\vp \simeq 0.7$.

\subsubsection{Case of varying $Q$}

While the above discussion corresponds to a constant value of 
$Q$ from the matter-dominant to the 
dilaton-dominant era, 
one can generalize the analysis to the case in which 
$Q$ is a time-varying function.
The reason why we did not get viable cosmological solutions
in the previous subsection is that the coupling between the matter 
and the dilaton is too large in the matter-dominant era. 
If $Q$ varies in a way such that it is small in the 
matter-dominant era but grows as the system enters
the dilaton-dominant phase, it is expected that 
we may get viable cosmological scaling solutions
at late times. 
The authors in Ref.~\cite{GPV} considered 
the coupling of the form
\begin{equation} 
\label{varyQ}
Q(\vp)=Q_0\frac{e^{Q_0\vp}}{b^2+e^{Q_0\vp}}\,,
\end{equation}
where $Q_0$ and $b$ are constants.
This is suppressed in the weakly coupled regime, but
approaches a constant value $Q_0$ as the field
evolves toward large values.
 
\begin{figure}
\epsfxsize = 4.0in \epsffile{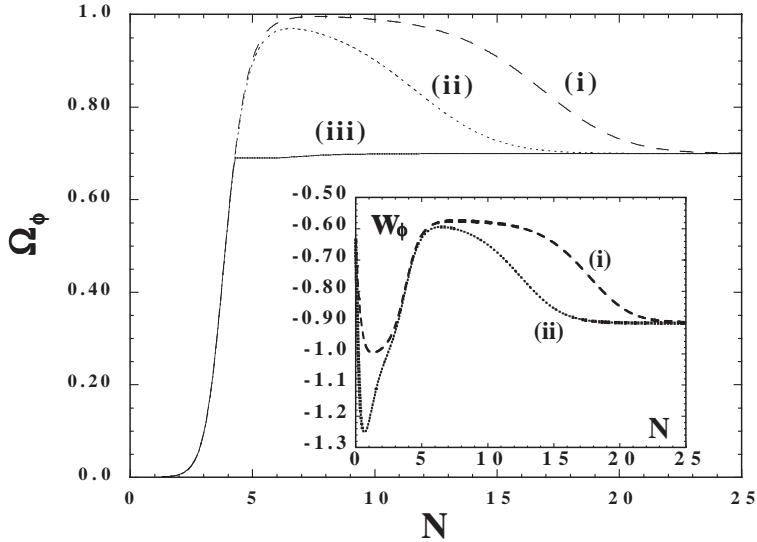} 
\caption{The variation of $\Omega_\vp$ in the case of 
the varying $Q$ for $\lambda=0.85$ and $c_1=c_2=1$
with initial conditions $x_i=0.003$, $z_i=0.0035$
and $\vp_i=6.6$.
The cases (i) and (ii) correspond to
the choices (i) $b=7 \times 10^3$, 
$Q_0=1.45$ and (ii) $b=2 \times 10^3$, 
$Q_0=1.45$ in Eq.~(\ref{varyQ}).
In the case (iii) we choose the coupling $Q$ which 
suddenly changes from $Q=0$ to $Q=Q_0$ 
around $\Omega_\vp=0.7$.
{\bf Inset}: The evolution of the equation of state 
$w_\vp$ for the cases (i) and (ii).}
\label{varyingq}
\end{figure}
  
In Fig.~\ref{varyingq} we plot the evolution of $\Omega_\vp$
for several different cases with $\lambda=0.85$, $c_1=1$
and $c_2=1$.
In the case (i), corresponding to $b=7 \times 10^3$ and
$Q_0=1.45$ in Eq.~(\ref{varyQ}), the coupling $Q$ is negligibly 
small during the matter-dominant era.
Therefore the dynamics of the system is similar to the 
$Q=0$ case before $\Omega_\vp$ reaches close to unity.
The effect of the coupling $Q$ becomes important after that,
which eventually leads to the approach to the scaling solutions
discussed in Sec.~\ref{sec_scaling}.
As seen in Fig.~\ref{varyingq} we numerically obtained attractor
values $\Omega_\vp=0.7$ and $w_\vp=-0.9$
for the choice $\lambda=0.85$ and $Q_0=1.45$.
Note that the equation of state always ranges
in the region $w_\vp>-1$, which means that the vacuum is 
stable at the quantum level. 
This case corresponds to the solution in which $\Omega_\vp$
is growing now ($N=4$) but approaches a constant value
with $0<\Omega_\vp<1$ in the future.
The asymptotic value of $\Omega_\vp$ is dependent on 
the choice of $\lambda$ and $Q_0$.

The initial values of $Q$ get larger if one chooses
smaller values of $b$ in Eq.~(\ref{varyQ}). 
Although this is required to make 
the time before the system approaches the scaling solution
shorter, the larger initial values of $Q$ lead to another 
problem. As seen in the inset of Fig.~\ref{varyingq},
the equation of state $w_\vp$ becomes less than $-1$
for the case (ii), marking the 
instability of the system. 
In order to avoid this problem, the initial value of $Q$
needs to be sufficiently small relative to 
unity as in the case (i) of Fig.~\ref{varyingq}.
If $Q$ jumps rapidly from nearly zero to the 
value $Q_0$ around $\Omega_\vp=0.7$,  it is possible 
to connect to the scaling solutions with $\Omega_\vp \simeq 0.7$
around $N=4$. This is shown in the case (iii) of Fig.~\ref{varyingq}.
in which we used the step-function for the coupling $Q$.

If one adopts the function (\ref{varyQ}), it is not easy to 
get the scaling solutions which {\it directly} approach 
$\Omega_\vp=0.7$
as in the case (iii) of Fig.~\ref{varyingq}.
The situation is similar when the potential of the dilaton 
is absent ($c_2=0$).
This is mainly because inclusion of the larger coupling $Q$ tends
to work to violate the stability condition (\ref{xi})
during the transition to the scalar field dominant phase.
Nevertheless it is intriguing that we found the scaling solutions
as in the type (i) that evolve to a constant $\Omega_\vp$
in the future without violating the   
stability condition.

\section{Conclusions}

In this paper we have studied a new type of dark energy
model based on string theory.
Our starting point is the Lagrangian (\ref{action})  
in low-energy effective string theory.
We adopted the runaway dilaton scenario in which 
the dilaton field $\phi$ is effectively decoupled from 
gravity as the field evolves toward $\phi \to \infty$.
The coefficient of the kinematic term of the dilaton can be negative 
in the Einstein frame, which means that the dilaton behaves as 
a phantom-type scalar field.
We implemented higher-order derivative 
terms for the field $\phi$ which ensure the stability of the system even when the 
coefficient of $\dot{\phi}^2$ is negative.

We then considered a general Lagrangian
which is generic function of a (canonically defined) 
scalar field $\vp$ and $X\equiv -(\nabla \varphi)^2/2$.
We first derived the stability condition of quantum 
fluctuations, see Eq.~(\ref{xi}).
At the classical level the perturbations may be stable if the 
speed of sound given in Eq.~(\ref{sound}) is positive, but we 
obtained a more stringent condition in order to ensure the 
stability of the vacuum at the quantum level. 
When the condition (\ref{xi})
is satisfied, one can avoid the  gravitational creation of 
ghost and photon pairs which was recently raised as a 
serious problem by Cline et al.\,\cite{Cline} for the 
classical phantom models considered 
so far \cite{caldwell,phantom,Sami:2003xv,sami}.
In this work we carefully addressed the way to 
overcome this problem.
The stability region is plotted in the phase-space,
which restricts the dynamical evolution of the system
relative to the normal scalar field with a positive 
kinematic term.
In Sec.\,\ref{sec_scaling} we derived a general form of the Lagrangian 
(\ref{scap}) for a single field $\vp$ when the scaling 
solutions exist.
We implemented a coupling between the dilaton and the matter
(denoted by $Q$) and obtained the equation of state $w_\vp$
and the relative energy density $\Omega_\vp$ for the 
field $\vp$, see Eq.~(\ref{womega}).
This is a general useful expression valid not only for the  
normal scalar field but also for the ghost-type 
scalar field.
We also found that when the Lagrangian is written as 
$p=f(X)-V(\phi)$ the requirement of the scaling solutions
uniquely determines the Lagrangian as
$p=\pm X-c_2e^{-\lambda \phi}$.
This is the system with an exponential potential and
a positive/negative kinematic term, which was  
extensively studied in the literature. 

We considered the classical dynamics of the system with a negative 
kinematic term $-X$ and the bell-type potential (\ref{po}), and reproduced the 
result that the dilaton evolves toward the potential {\it maximum}
with an equation of state $w_\vp \le -1$.
However, since this system is unstable at the quantum level, we took
into account a term of the type $e^{\lambda \vp}X^2$ in order to ensure
the stability of quantum fluctuations. When the condition for the 
stability is satisfied, the equation of state is restricted in the range
$w_\vp \ge -1$ in general.
We found that there exist attractor solutions with 
$Q=0$ in which $\Omega_\vp$ evolves toward 1 and 
$w_\vp$ approaches a constant value slightly larger than $-1$. 
This is different from the scaling solutions discussed in 
Sec.\,\ref{sec_scaling} which keep the relative energy density constant 
with {\it nonzero} values of $\Omega_\vp$ and $\Omega_m$.
This can be derived by setting $\rd x/\rd N=\rd y/\rd N=0$ 
in the evolution equations (\ref{xeq}) and (\ref{zeq}).
We plotted such examples in Figs.\,~\ref{evon} and \ref{evo},
whose results agree well with analytic estimates.
We also checked that the system lies in the quantum stability
region shown in Fig.~\ref{stability} throughout the evolution, 
thereby providing a viable dark energy model
even when the coefficient of the $\dot{\vp}^2$ term is negative. 
When the matter Lagrangian $\wt{{\cal S}}_m$ is not dependent on the field 
$\phi$ in the string frame, one can show that $Q \propto (1/B_g)
\rd B_g/\rd \phi$ is vanishing 
as $\phi \to \infty$ for the choice (\ref{triv}).
In this case one can get ideal cosmological solutions 
as shown in Figs.~\ref{evon} and \ref{evo}.
If the matter is coupled to the dilaton in the string frame, the coupling 
$Q$ is not necessarily zero. We analytically found in Sec.\,\ref{sec_scaling}
that scaling solutions exist when $Q$ is a nonzero constant
in the scaling regime.
When $Q$ is a constant from the matter-dominated to the dilaton-dominated
era, the stability condition is violated for the realistic initial conditions
($x_i^2, z_i^2 \ll 1$). Therefore it is difficult to obtain a viable cosmological 
evolution in such a case, see Figs.~\ref{pspace} and \ref{phase}.
However, if the coupling $Q$ rapidly grows from nearly zero 
to a constant value as in Eq.~(\ref{varyQ}), the system can approach
the scaling solution without breaking the stability condition, 
see the case (i) of Fig.~\ref{varyingq}.
In order to get a ``direct'' approach toward $\Omega_\vp \simeq 0.7$
as in the case (iii) of Fig.~\ref{varyingq},
we require a step-like change of the coupling $Q$ during the transition 
to the scalar-field dominant era.

In our work we carried out the analysis for the simplified Lagrangian 
(\ref{ourlag}) in order to understand the basic picture of the system 
with a phantom-type scalar field.
In string theory we have other non-perturbative and loop corrections 
such as the Gauss-Bonnet curvature invariant \cite{Cartier,Tsuji,Foffa}.
It is certainly of interest to extend our analysis
to such a direction.

\section*{ACKNOWLEDGMENTS}
We are grateful to Parampreet Singh for his contribution at the early 
stages of this work and for very useful discussions and comments
on the draft.
It is also a pleasure to thank Luca Amendola, Maurizio Gasperini, 
Luciano Girardello, 
Enrico Trincherini, Gabriele Veneziano and Alberto Zaffaroni 
for very useful discussions. The research of F.P. is partially supported by
MURST (contract 2003-023852-008) and by the European Commision (TMR 
program HPRN-CT-2000-00131).
The research of S.T. is supported by a grant 
from JSPS (No.\,04942). 
F.P. thanks CERN, and 
S.T. thanks  IUCAA and Rome observatory for supporting 
their visits and  warm hospitality 
during various stages of completion of this work.

\section*{APPENDIX : Classical dynamics of the phantom field}\label{classi}

In this Appendix we consider the 
classical evolution of the system in the 
absence of  the higher-order terms in $X$
with the exponential potential of the dilaton 
($c_1=0$, $c_2 \ne 0$ and $Q=0$).
Although the stability of quantum fluctuations 
is not ensured for $c_1=0$, it is worth clarifying
the classical evolution to compare it 
with the case in which higher-order terms are included.
When $Q=0$ we obtain the following fixed points
by setting $\rd x/\rd N=0$ and $\rd z/\rd N=0$ in 
Eqs.~(\ref{xeq}) and (\ref{zeq}):
\begin{equation} 
\label{fixed}
(x, z)=(0, 0),~~\left(-\frac{\lambda}{\sqrt{6}}, \sqrt{\frac{1}{c_2}
\left(1+\frac{\lambda^2}{6}\right)}\right)\,.
\end{equation}
The first solution is a trivial one which approaches 
$\Omega_\vp \to 0$ at late times.
On the other hand the second one corresponds to 
an asymptotic solution with $\Omega_\vp \to 1$ at late times.
This differs from the scaling solutions that approach
a nonzero value of $\Omega_\vp$ or $\Omega_m$.
The scaling solutions which keep 
$\Omega_\vp$ constant in the range $0<\Omega_\vp<1$ do not 
exist in the present case, since $\Omega_\vp$ becomes negative 
in Eq.~(\ref{likebefore}) for $Q=0$ and $\epsilon=-1$.

{}When $c_1=0$ the equation of state for the field $\vp$ is given as
\begin{equation} 
\label{wphi2}
w_\vp=\frac{-Y-c_2}{-Y+c_2}\,.
\end{equation}
{}From Eq.~(\ref{quan}) the function $Y=Xe^{\lambda\vp}$
is generally expressed as
\begin{equation} 
\label{Ygeneral}
Y=x^2/z^2\,.
\end{equation}
Therefore the second fixed point in Eq.~(\ref{fixed})
corresponds to the equation of state
\begin{equation} 
\label{exp}
w_\vp=-1-\lambda^2/3\,,
\end{equation}
which is less than $-1$. 
It is clear that this is different from the scaling solution
(\ref{likebefore}) with $Q=0$, since $w_{\varphi} \to 0$ in this case.
This constant behavior of $w_\vp$ for the exponential potential 
was numerically found in Ref.~\cite{Sami:2003xv}, but we succeeded
to derive it analytically.
The field $\vp$ is driven up the potential hill due to the 
negative kinematic term.

We can consider the bell-type potential (\ref{po}) in order to follow
the classical evolution of the dilaton. Then the dilaton evolves 
toward the potential {\it maximum} and 
is eventually stabilized there.
In the region where the potential (\ref{po}) is effectively 
described by an exponential one, we have numerically checked
the existence of the constant equation of state characterized 
by Eq.~(\ref{exp}).
For the bell-type potential, this stage is only transient and 
$w_\vp$ approaches $-1$ as the field evolves
toward the potential maximum. 
After the field is stabilized there, we have $x=0$ and $Y=0$,
thus yielding $w_\vp=-1$.
This classical evolution of the dilaton is similar to what was
already discussed in Refs.~\cite{Carroll,sami}, hence 
we do not repeat it here.
If one takes into account the coupling between the dilaton and 
the matter ($Q \ne 0)$, we have additional scaling solutions 
characterized by Eq.~(\ref{likebefore}).



\begin{thebibliography}{99}

\bibitem{varun}
V.~Sahni,
Class.\ Quant.\ Grav.\  {\bf 19}, 3435 (2002)
[arXiv:astro-ph/0202076];
[arXiv:astro-ph/0403324].

\bibitem{paddy}
T.~Padmanabhan,
Phys.\ Rept.\  {\bf 380}, 235 (2003)
[arXiv:hep-th/0212290].



\bibitem{quin}
R.~R.~Caldwell, R.~Dave and P.~J.~Steinhardt,
Phys.\ Rev.\ Lett.\  {\bf 80}, 1582 (1998)
[arXiv:astro-ph/9708069];
I.~Zlatev, L.~M.~Wang and P.~J.~Steinhardt,
Phys.\ Rev.\ Lett.\  {\bf 82}, 896 (1999)
[arXiv:astro-ph/9807002].

\bibitem{kesse}
C.~Armendariz-Picon, V.~Mukhanov and P.~J.~Steinhardt,
Phys.\ Rev.\ Lett.\  {\bf 85}, 4438 (2000)
[arXiv:astro-ph/0004134];
C.~Armendariz-Picon, V.~Mukhanov and P.~J.~Steinhardt,
Phys.\ Rev.\ D {\bf 63}, 103510 (2001)
[arXiv:astro-ph/0006373];
T.~Chiba, T.~Okabe and M.~Yamaguchi, 
Phys.\ Rev.\ D {\bf 62}, 023511 (2000)
[arXiv:hep-ph/9907402].

\bibitem{brane}
G.~R.~Dvali, G.~Gabadadze and M.~Porrati,
Phys.\ Lett.\ B {\bf 485}, 208 (2000)
[arXiv:hep-th/0005016].

\bibitem{tachyon}
A.~Sen,
JHEP {\bf 0207}, 065 (2002)
[arXiv:hep-th/0203265];
T.~Padmanabhan,
Phys.\ Rev.\ D {\bf 66}, 021301 (2002)
[arXiv:hep-th/0204150].

\bibitem{chap}
M.~C.~Bento, O.~Bertolami and A.~A.~Sen,
Phys.\ Rev.\ D {\bf 66}, 043507 (2002)
[arXiv:gr-qc/0202064].

\bibitem{TV88} 
T.~R.~Taylor and G.~Veneziano, Phys.\ Lett.\ B {\bf 213}, 450 (1988).

\bibitem{DP94} T. ~Damour and A.~M.~Polyakov, Nucl.\ Phys.\ 
{\bf B423}, 532 (1994);
Gen.\ Rel.\ Grav. {\bf 26}, 1171 (1994).

\bibitem{V01} G.~Veneziano, J.\ High\ Energy\ Phys.\ {\bf 06}, 051 (2002)
[arXiv:hep-th/0110129].

\bibitem{GPV}
M.~Gasperini, F.~Piazza and G.~Veneziano,
Phys.\ Rev.\ D {\bf 65}, 023508 (2002)
[arXiv:gr-qc/0108016].

\bibitem{DPV1}
T. ~Damour, F. ~Piazza and G. ~Veneziano, 
Phys.\ Rev.\ Lett. {\bf 89} 81601 (2002) 
[arXiv:gr-qc/0204094].

\bibitem{DPV2}
T.~Damour, F.~Piazza and G.~Veneziano,
Phys.\ Rev.\ D {\bf 66}, 046007 (2002)
[arXiv:hep-th/0205111].

\bibitem{wetter}
C.~Wetterich, Astron. Astrophys. {\bf 301}, 321 (1995)
[arXiv:hep-th/9408025].

\bibitem{Amendola2}
L.~Amendola,
Phys.\ Rev.\ D {\bf 62}, 043511 (2000)
[arXiv:astro-ph/9908023].

\bibitem{AGUT} 
L.~Amendola, M.~Gasperini, 
D. Tocchini-Valentini and C. Ungarelli, 
Phys.\ Rev.\ D {\bf 67},  043512 (2003)
[arXiv:astro-ph/0208032].

\bibitem{Amendola3} L. Amendola, 
Mon.\ Not.\ Roy.\ Astron.\ Soc. {\bf 342} 221 (2003)
[arXiv:astro-ph/0209494].

\bibitem{AGPV}
 L.~Amendola, M.~Gasperini and F.~Piazza, in 
preparation.

\bibitem{caldwell}
R.~R.~Caldwell,
Phys.\ Lett.\ B {\bf 545}, 23 (2002)
[arXiv:astro-ph/9908168].

\bibitem{tirthalam} 
A.~G.~Riess et al. arXiv:astro-ph/0402512;
T.~R.~Choudhury and T.~Padmanabhan,
arXiv:astro-ph/0311622;
U. Alam et al., arXiv:astro-ph/0311364.

\bibitem{phantom}
A.~E.~Schulz and M.~J.~White,
Phys.\ Rev.\ D {\bf 64}, 043514 (2001)
[arXiv:astro-ph/0104112];
R.~R.~Caldwell, M.~Kamionkowski and N.~N.~Weinberg,
Phys.\ Rev.\ Lett.\  {\bf 91}, 071301 (2003)
[arXiv:astro-ph/0302506];
G.~W.~Gibbons,
arXiv:hep-th/0302199;
S.~Nojiri and S.~D.~Odintsov,
Phys.\ Lett.\ B {\bf 562}, 147 (2003)
[arXiv:hep-th/0303117];
Phys.\ Lett.\ B {\bf 571}, 1 (2003)
[arXiv:hep-th/0306212];
P.~F.~Gonzalez-Diaz,
Phys.\ Rev.\ D {\bf 68}, 021303 (2003)
[arXiv:astro-ph/0305559];
J.~g.~Hao and X.~z.~Li,
Phys.\ Rev.\ D {\bf 68}, 043501 (2003)
[arXiv:hep-th/0305207];
arXiv:astro-ph/0309746;
M.~P.~Dabrowski, T.~Stachowiak and M.~Szydlowski,
Phys.\ Rev.\ D {\bf 68}, 103519 (2003)
[arXiv:hep-th/0307128];
L.~P.~Chimento and R.~Lazkoz,
Phys.\ Rev.\ Lett.\  {\bf 91}, 211301 (2003)
[arXiv:gr-qc/0307111];
D.~j.~Liu and X.~z.~Li,
Phys.\ Rev.\ D {\bf 68}, 067301 (2003)
[arXiv:hep-th/0307239];
H.~Stefancic,
arXiv:astro-ph/0310904;
E.~Elizalde and J.~Q.~Hurtado,
Mod.\ Phys.\ Lett.\ A {\bf 19}, 29 (2004)
[arXiv:gr-qc/0310128];
V.~B.~Johri,
arXiv:astro-ph/0311293;
H.~Q.~Lu,
arXiv:hep-th/0312082;
H.~Stefancic,
arXiv:astro-ph/0312484;
P.~F.~Gonzalez-Diaz,
Phys.\ Lett.\ B {\bf 586}, 1 (2004)
[arXiv:astro-ph/0312579];
M.~Szydlowski, W.~Czaja and A.~Krawiec,
arXiv:astro-ph/0401293;
J.~M.~Aguirregabiria, L.~P.~Chimento and R.~Lazkoz,
arXiv:astro-ph/0403157;
M.~Bouhmadi-Lopez and P.~Vargas Moniz,
arXiv:gr-qc/0404111;
Z.~K.~Guo, Y.~S.~Piao and Y.~Z.~Zhang,
arXiv:astro-ph/0404225.

\bibitem{Sami:2003xv}
M.~Sami and A.~Toporensky,
arXiv:gr-qc/0312009.

\bibitem{sami}
P.~Singh, M.~Sami and N.~Dadhich,
Phys.\ Rev.\ D {\bf 68}, 023522 (2003)
[arXiv:hep-th/0305110].

\bibitem{Carroll}
S.~M.~Carroll, M.~Hoffman and M.~Trodden,
Phys.\ Rev.\ D {\bf 68}, 023509 (2003)
[arXiv:astro-ph/0301273].

\bibitem{Cline}
J.~M.~Cline, S.~y.~Jeon and G.~D.~Moore,
arXiv:hep-ph/0311312.

\bibitem{Arkani}
N.~Arkani-Hamed, H.~C.~Cheng, M.~A.~Luty and S.~Mukohyama,
arXiv:hep-th/0312099.

\bibitem{kinflation}
C.~Armendariz-Picon, T.~Damour and V.~Mukhanov,
Phys.\ Lett.\ B {\bf 458}, 209 (1999)
[arXiv:hep-th/9904075].

\bibitem{Arkani2}
N.~Arkani-Hamed, P.~Creminelli, S.~Mukohyama and M.~Zaldarriaga,
JCAP {\bf 0404}, 001 (2004)
[arXiv:hep-th/0312100].

\bibitem{ghostrecent}
S.~L.~Dubovsky,
arXiv:hep-ph/0403308;
M.~Peloso and L.~Sorbo,
arXiv:hep-th/0404005;
A.~V.~Frolov,
arXiv:hep-th/0404216;
L.~Senatore,
arXiv:astro-ph/0406187;
V.~V.~Kiselev,
arXiv:gr-qc/0406086.

\bibitem{Green}
M.~B.~Green, J.~Schwartz and E.~Witten, 
{\it Superstring Theory} (Cambrigde University 
Press, Cambrigde, England, 1987).

\bibitem{Cartier}
C.~Cartier, J.~c.~Hwang and E.~J.~Copeland,
Phys.\ Rev.\ D {\bf 64}, 103504 (2001)
[arXiv:astro-ph/0106197].

\bibitem{Tsuji}
S.~Tsujikawa, R.~Brandenberger and F.~Finelli,
Phys.\ Rev.\ D {\bf 66}, 083513 (2002)
[arXiv:hep-th/0207228];
S.~Tsujikawa,
Phys.\ Lett.\ B {\bf 526}, 179 (2002)
[arXiv:gr-qc/0110124];
Class.\ Quant.\ Grav.\  {\bf 20}, 1991 (2003)
[arXiv:hep-th/0302181].


\bibitem{Cope}
E.~J.~Copeland, A.~R.~Liddle and D.~Wands,
Phys.\ Rev.\ D {\bf 57}, 4686 (1998) 
[arXiv:gr-qc/9711068] (5-th solution in Table I); 
P.~G.~Ferreira and M.~Joyce Phys. Rev.\ D {\bf 58}, 023503 (1998)
[arXiv:astro-ph/9711102].

\bibitem{Kos}
V.~A.~Kostelecky, R.~Lehnert and M.~J.~Perry,
Phys.\ Rev.\ D {\bf 68}, 123511 (2003);
O.~Bertolami, R.~Lehnert, R.~Potting, and A.~Ribeiro, 
Phys.\ Rev.\ D {\bf 69}, 083513 (2004) [arXiv:astro-ph/0310344].

\bibitem{kper}
J.~Garriga and V.~F.~Mukhanov,
Phys.\ Lett.\ B {\bf 458}, 219 (1999)
[arXiv:hep-th/9904176].

\bibitem{Foffa}
S.~Foffa, M.~Maggiore and R.~Sturani,
Nucl.\ Phys.\ B {\bf 552}, 395 (1999)
[arXiv:hep-th/9903008].


\end{thebibliography}
\end{document}